\documentclass[10pt, conference]{IEEEtran}
\usepackage{mathptmx}
\usepackage{fancyhdr}
\usepackage[normalem]{ulem}
\usepackage[hyphens]{url}
\usepackage[sort,nocompress]{cite}
\usepackage[final]{microtype}
\usepackage[keeplastbox]{flushend}
\usepackage[bookmarks=true,breaklinks=true,letterpaper=true,colorlinks,citecolor=blue,linkcolor=blue,urlcolor=blue]{hyperref}
\usepackage{tikz}
\usepackage{xcolor}
\usepackage[normalem]{ulem}
\usepackage{amsmath}
\usepackage{multirow}
\usepackage{diagbox}
\usepackage{makecell}
\usepackage[linesnumbered,ruled,vlined]{algorithm2e}
\usepackage[capitalise]{cleveref}

\newcommand{\frameworkname}{SymBreak}

\usepackage{authblk}
\usepackage{titlesec}
\titleformat{\title}{\normalfont\LARGE\bfseries}{}{}{}
\title{\frameworkname: Mitigating Quantum Degeneracy Issues in QLDPC Code Decoders by Breaking Symmetry}

\author[1]{Keyi Yin\thanks{keyin@ucsd.edu}}
\author[1]{Xiang Fang\thanks{x8fang@ucsd.edu}}
\author[1]{Jixuan Ruan\thanks{j3ruan@ucsd.edu}}
\author[1]{Hezi Zhang\thanks{hez019@ucsd.edu}}
\author[1]{Dean Tullsen\thanks{tullsen@ucsd.edu}} 
\author[2]{\authorcr Andrew Sornborger\thanks{sornborg@lanl.gov}}
\author[3]{Chenxu Liu\thanks{chenxu.liu@pnnl.gov}}
\author[3]{Ang Li\thanks{ang.li@pnnl.gov}}
\author[4]{Travis Humble\thanks{humblets@ornl.gov}}
\author[1]{Yufei Ding\thanks{yufeiding@ucsd.edu}}

\affil[1]{University of California, San Diego, CA, USA}
\affil[2]{Los Alamos National Laboratory, Los Alamos, NM, USA}
\affil[3]{Pacific Northwest National Laboratory, Richland, WA, USA}
\affil[4]{Oak Ridge National Laboratory, Oak Ridge, TN, USA}

\begin{document}
\maketitle
\pagestyle{plain}
\begin{abstract}
Quantum error correction (QEC) is critical for scalable and reliable quantum computing, but existing solutions, such as surface codes, incur significant qubit overhead. Quantum low-density parity check (qLDPC) codes have recently emerged as a promising alternative, requiring fewer qubits. However, the lack of efficient decoders remains a major barrier to their practical implementation. In this work, we introduce \frameworkname, a novel decoder for qLDPC codes that adaptively modifies the decoding graph to improve the performance of state-of-the-art belief propagation (BP) decoders. Our key contribution is identifying quantum degeneracy as a root cause of the convergence issues often encountered in BP decoding of quantum LDPC codes. We propose a solution that mitigates this issue at the decoding graph level, achieving both fast and accurate decoding. Our results demonstrate that \frameworkname outperforms BP and BP+OSD—a more complex variant of BP—with a 16.17× reduction in logical error rate compared to BP and 3.23× compared to BP+OSD across various qLDPC code families. With only an 18.97\% time overhead compared to BP, \frameworkname provides significantly faster decoding times than BP+OSD, representing a major advancement in efficient and accurate decoding for qLDPC-based QEC architectures.
\end{abstract}

\section{Introduction}
\label{sec:introduction}
Quantum error correction (QEC)~\cite{nielsen2010quantum, knill1997theory, gottesman2010introduction, shor1996fault} is essential for overcoming \emph{noise}, the primary barrier to building reliable and scalable quantum computers for practical applications~\cite{shor1999polynomial, grover1996fast, cao2019quantum}. This technique encodes quantum information into \emph{QEC codes}~\cite{calderbank1996good, steane1996multiple, bombin2006topological, kitaev2003fault, fowler2012surface} using redundant qubits to gain resilience to errors. During the computation, a QEC protocol consistently detects and corrects the errors occurred, ensuring the robustness of the quantum system. 

\begin{figure}[ht]
\centering
\includegraphics[scale=0.17]{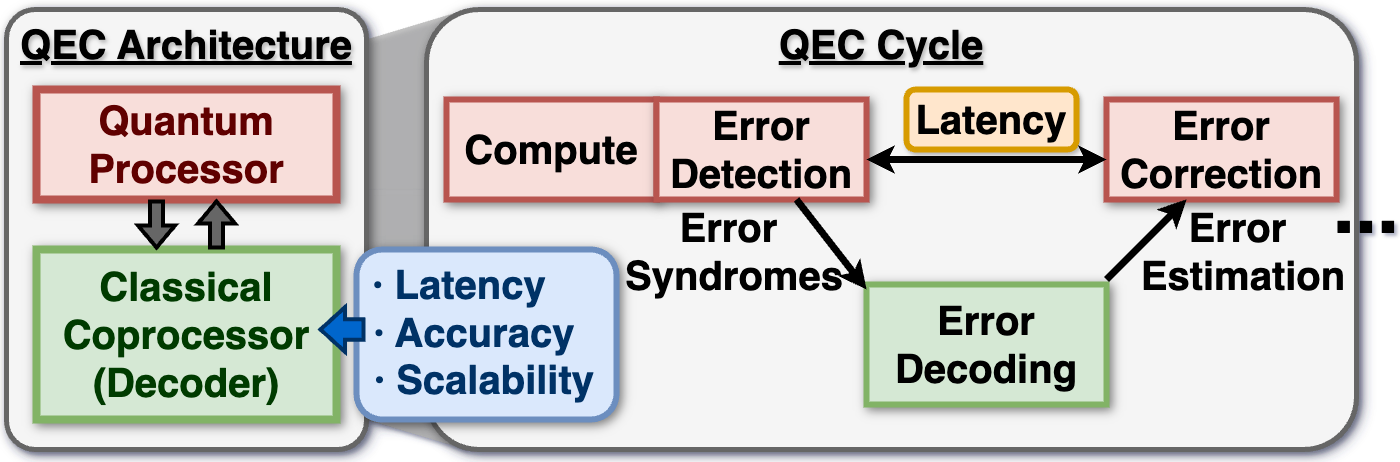}
\caption{QEC Architecture.}
\label{fig: QEC Arch}
\end{figure}

Implementing QEC requires an architecture that integrates a \emph{quantum processor} with a \emph{classical coprocessor} (Fig.~\ref{fig: QEC Arch}). In each QEC cycle, the quantum processor executes computations and detects errors through measurements, producing \emph{error syndromes}. These syndromes are sent to the classical coprocessor, known as the \emph{decoder}, which identifies errors and sends corrective actions back to the quantum processor (Fig.~\ref{fig: QEC Arch}). By iterating these QEC cycles, the architecture performs computation while protecting against noise.

This architecture imposes stringent requirements on the decoder in three key aspects: (1) \emph{Complexity.} The decoder must operate in \emph{real-time} with low complexity to keep up with rapid quantum operations, which can require a few microseconds per QEC cycle on some quantum hardware~\cite{acharya2024quantum}.
(2) \emph{Accuracy.} The decoder must provide precise error corrections to prevent uncorrectable error accumulation. (3) \emph{Scalability.} The decoder must handle error syndromes efficiently for large, practical quantum systems. Developing a high-performance decoder that meets these requirements is challenging but essential for the effectiveness of the QEC architecture.

In the vein of QEC architectures, \emph{Surface-codes}~\cite{bravyi1998quantum, dennis2002topological, litinski2019game, beverland2022surface} have been leading solutions due to their simple 2D grid structures and efficient MWPM (Minimum Weight Perfect Matching) decoders~\cite{delfosse2021almost, delfosse2020linear, higgott2022pymatching, alavisamani2024promatch, vittal2023astrea, das2022afs}, with successful demonstrations in physical experiments~\cite{google2023suppressing, acharya2024quantum}. However, surface codes incur large qubit overhead that grows \emph{quadratically} with the number of correctable errors, with practical applications estimated to need millions of qubits~\cite{gidney2021factor}. Recently, \emph{quantum low-density parity check} (qLDPC) codes~\cite{gottesman2013fault, tillich2013quantum, leverrier2015quantum, panteleev2022asymptotically, bravyi2024high, leverrier2022quantum, dinur2023good} have gained attention for their \emph{linear} qubit overhead, allowing them to store equivalent quantum information with the same level of protection using significantly fewer qubits than surface codes~\cite{bravyi2024high, xu2024constant, scruby2024high} and offering a promising path toward practical, large-scale quantum computation. This advantage arises from their complex structures, which involve non-local and high-degree connectivity~\cite{bravyi2009no, bravyi2010tradeoffs}. These features not only complicate hardware implementation but also render effective MWPM decoders~\cite{alavisamani2024promatch, vittal2023astrea, das2022afs}, which work effectively and efficiently for surface codes, inapplicable~\cite{delfosse2021almost, higgott2022pymatching, demartireview}. 

\begin{figure*}[!ht]
    \centering
    \includegraphics[width=0.95\textwidth]{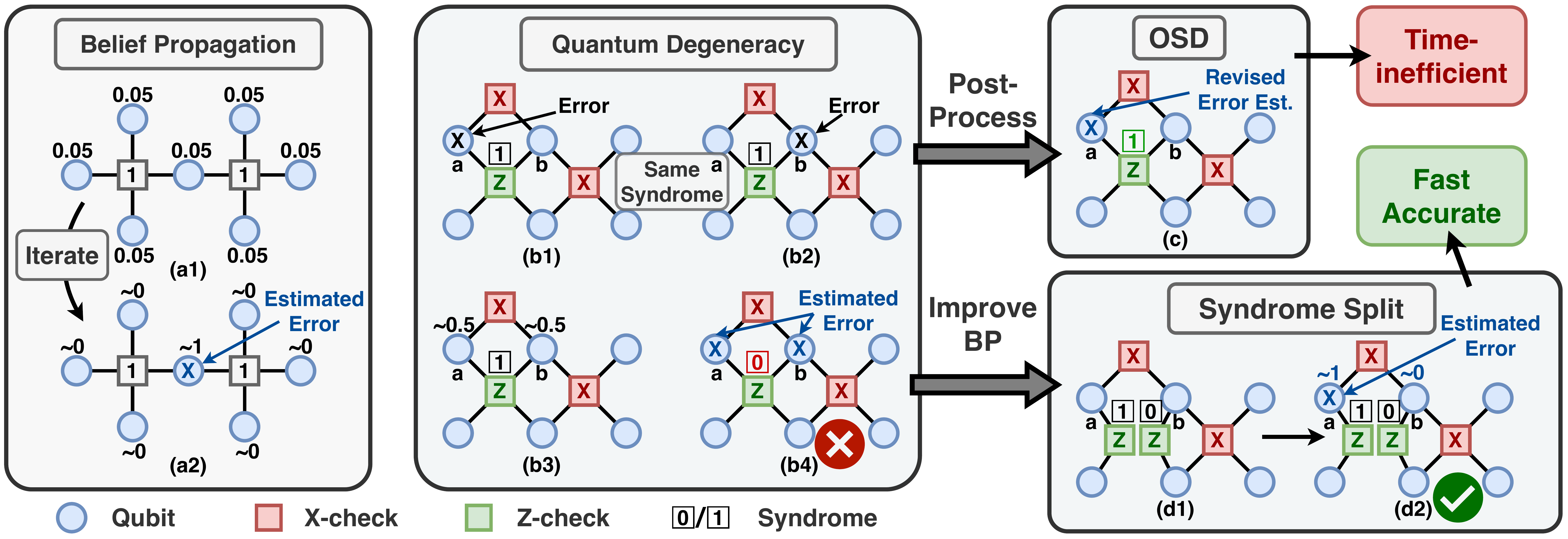}
    \caption{(a1)(a2) Mechanism of BP: updating error probabilities to obtain error estimations. (b1)(b2) Example of quantum degeneracy: two distinct errors correspond to the same syndrome. (b3) BP assigns equal probability to both errors. (b4) BP incorrectly assumes both qubits have errors, resulting in an error estimation that does not match the syndrome. (c) OSD provides a revised error estimation that matches the syndrome but is time-inefficient. (d) Syndrome split enables BP to converge to a correct error estimation efficiently and accurately.}
    \label{fig:intro}
     \vspace{-2pt}
\end{figure*}

Recent hardware advancements  have laid a promising solution for implementing qLDPC codes. Neutral atom platforms~\cite{bluvstein2024logical, bluvstein2022quantum}, with their ability to move qubits, naturally support non-local connectivity, providing a clear advantage for exploring qLDPC codes. 
Superconducting qubit platforms are also making progress. IBM has proposed a practical approach by investigating simpler non-local structures to facilitate efficient implementation of qLDPC codes~\cite{tremblay2022constant}. For example, a new family of qLDPC codes, called \emph{bivariate bicycle} (BB) codes~\cite{bravyi2024high}, offers favorable properties such as low qubit overhead, high resilience to errors, a relatively low qubit degree of 6, and a manageable 2-layer structure. These advancements position BB codes as a promising candidate for low-overhead QEC demonstrations with current quantum technology.

With hardware implementation challenges largely addressed, the lack of efficient and effective decoders has emerged as the primary bottleneck for qLDPC-based QEC architectures. State-of-the-art qLDPC decoders  attempt to address this by combing \emph{belief propagation} with post-processing via \emph{order statistics decoding} (BP + OSD)~\cite{roffe2020decoding, panteleev2021degenerate}, but significant limitations persist.  The BP step iteratively updates estimated \emph{error probability} (EP) for each qubit based on error syndromes and determines error presence based on whether the EP is above or below $0.5$ (see Fig.~\ref{fig:intro}(a1)(a2)). This step is efficient with time complexity of $O(n)$ (where $n$ is the number of qubits)~\cite{mackay1997near}. However, the error estimations from BP may not align with the observed error syndromes. The OSD step mitigates this by adjusting the least reliable estimates (those with EPs close to 0.5) to match the syndromes. Despite its accuracy, OSD is too complicated for real-time implementations, failing even for codes with $\sim$100 qubits~\cite{valls2021syndrome}, let alone larger codes necessary for useful quantum applications.

To overcome the limitations of the BP+OSD decoder, it is essential to understand why BP fails to provide an accurate error estimation.
We identify the primary cause as \emph{quantum degeneracy}, where distinct error patterns produce identical error syndromes—a common phenomenon in QEC codes~\cite{roffe2020decoding, panteleev2021degenerate, fuentes2021degeneracy}. BP tends to assign equal probability to all such patterns, often resulting in incorrect error estimations. For instance, in Fig.\ref{fig:intro}(b1)(b2),  $X$-errors on qubits $a$ and $b$ both yield the same syndrome of $1$ on the $Z$-check. As BP iterates, the EPs for qubits $a$ and $b$ converge to $0.5$ (Fig.~\ref{fig:intro}(b3)). If these EPs are just slightly above $0.5$, BP will assume both qubits have errors; however, this error estimation results in an incorrect syndrome of $1 \oplus 1 = 0$ and cannot be accepted (Fig.~\ref{fig:intro}(b4)), thus triggering the OSD step. Although OSD can revise the error estimation to match the syndrome, it is generally very time-inefficient.

To address these drawbacks, we propose a novel decoder, \textbf{\frameworkname}, which effectively tackles the degeneracy issue, enhancing BP performance and potentially eliminating the need for the complex OSD step. 
Unlike standard BP, which updates EPs on a static decoding graph where degeneracy leads to unreliable error estimations, \frameworkname~adaptively modifies the decoding graph based on information gathered during BP, guiding EP updates to converge toward 0 or 1 instead of 0.5, improving the the reliability of error estimation. 

While conceptually straightforward, modifying the graph to improve BP performance requires careful handling. This modification must (1) precisely disrupt the degeneracy structure, (2) retain as much error information as possible, and (3) be fast. 
To address these challenges, we developed a \emph{syndrome split} technique, illustrated in Fig.\ref{fig:intro}(d1)(d2). First, we identify the syndrome node (squares in Fig.~\ref{fig:intro}) most likely affected by quantum degeneracy based on a metric derived from BP’s intermediate outputs. We then split this syndrome node into two, strategically distributing the qubits connected to the original node between the two new nodes to disrupt the underlying degeneracy structure. Finally, we assign reliable estimates to the two new syndromes to support convergence in subsequent BP rounds. This approach ``breaks'' the degeneracy structure without removing nodes or edges, thus preserving the error information to the best extent. Importantly, all these operations are simple and time-efficient, enabling a fast, linear BP-based decoder capable of handling large-scale codes with thousands of qubits while maintaining good accuracy. Detailed discussions are provided in Sec.~\ref{sec: break dengen} and Sec.~\ref{sec: check split}. In particular, we identify an alternative approach for classifying qubits within the BB codes into distinct layers, providing an intuitive method for syndrome splitting that enables their fast and accurate decoding (Sec.~\ref{sec: BB code decoder}).

\vspace{5pt}
To summarize, the contributions of this paper are:
\vspace{-5pt}
\begin{itemize}
\item We propose a novel decoder, \textbf{\frameworkname}, for qLDPC codes, achieving both low decoding complexity and high accuracy.

\item We introduce syndrome split, a method that effectively addresses quantum degeneracy in qLDPC codes, enhancing BP performance and enabling an efficient decoder.


\item Our evaluation shows that \frameworkname~eliminates OSD-related steps while reducing the logical error rate by 16.17× compared to BP and 3.23× compared to BP+OSD across various qLDPC codes. Additionally, it incurs only a mild 18.97\% time overhead relative to BP, while offering orders-of-magnitude time savings over BP+OSD.
\end{itemize}





\section{Background}
\label{sec:background}
This section begins with an introduction to QEC basics and qLDPC codes, with a focus on the BB code and its favorable properties, followed by an overview of the BP+OSD decoder.

\subsection{QEC Basics}
\vspace{3pt}
\noindent\textbf{QEC Codes.} An $[[n,k,d]]$ QEC code encodes $k$ logical qubits into $n$ physical qubits, with an encoding rate of $k/n$, where a higher rate implies fewer physical qubits needed to store the same quantum information. The code includes $n-k$ \emph{parity checks} for error detection, each represented by a Pauli operator~\cite{nielsen2010quantum} acting on multiple qubits. To encode nontrivial information, all parity checks must commute. During each QEC cycle, these parity checks are measured to produce \emph{error syndromes} valued at 0 or 1, which are subsequently decoded to identify and correct errors. Some errors, however, alter the quantum state in a way that parity checks cannot detect, resulting in a \emph{logical error}. The smallest number of single-qubit errors needed to produce a logical error is called the \emph{code distance} $d$, where a higher code distance indicates greater error tolerance. A desirable QEC code should have a high encoding rate and a large distance relative to the code size $n$.


\vspace{3pt}
\noindent\textbf{Tanner Graph.} The structure of a QEC code can be represented by a Tanner graph, which consists of two types of nodes: \emph{variable nodes} and \emph{check nodes} (or \emph{syndrome nodes}), corresponding to qubits and parity checks, depicted as circles and squares in Fig.~\ref{fig: BB}, respectively. An edge connects a check node to a variable node if the associated parity check involves that qubit. This graph-based representation enables the application of classical graph theory tools, providing a foundation for reasoning about the design of decoders. 

\vspace{3pt}
\noindent\textbf{Decoder.} A QEC decoder is a classical algorithm that processes the error syndromes from parity checks and estimates the most probable error pattern that matches the observed syndromes. The MWPM decoder~\cite{dennis2002topological, delfosse2020linear, higgott2022pymatching}, widely used for surface codes~\cite{fowler2012surface}, has made significant advances in efficiency and accuracy over the past two decades. It reformulates the decoding problem as a standard weighted matching problem~\cite{livi2013graph} on a decoding graph derived from the Tanner graph. In this transformation, parity checks are treated as nodes and qubits as edges, with edge weights assigned based on error rates. For the matching problem to be well-defined, each qubit must connect to exactly two parity checks, allowing it to correspond to a single edge in the decoding graph. However, if a qubit connects to more than two parity checks, it cannot be represented as a simple edge and requires a more complex structure, which falls outside the scope of standard matching algorithms.
This restriction prevents the application of MWPM decoders to more general QEC codes with higher connectivity degrees, such as qLDPC codes (Sec.~\ref{subsec: qLDPC BB}), which often have superior properties.


\subsection{qLDPC Codes, Bivariate Bicycle (BB) Code}{\label{subsec: qLDPC BB}}

\noindent\textbf{qLDPC Codes.} qLDPC codes have bounded connectivity degrees for both variable and check nodes~\cite{gottesman2013fault}. These connections can be geometrically non-local and extend beyond 2D~\cite{bravyi2009no, bravyi2010tradeoffs}, allowing for a favorable constant asymptotic encoding rate as the code scales~\cite{tillich2013quantum, leverrier2015quantum, panteleev2022asymptotically, leverrier2022quantum, dinur2023good}, resulting in substantial qubit savings. The higher connectivity in qLDPC codes not only presents challenges for physical implementation but also precludes the use of efficient MWPM decoders commonly applied to surface codes~\cite{delfosse2021almost, delfosse2020linear, higgott2022pymatching}. While recent advances in quantum hardware with non-local interactions have made qLDPC codes more feasible to implement~\cite{bluvstein2024logical,tremblay2022constant} and several decoders have been developed~\cite{leverrier2015quantum, fawzi2018efficient, panteleev2021degenerate, roffe2020decoding, delfosse2022toward}, constructing qLDPC codes with relatively simple structures and strong properties and developing novel decoding methods for them remains an active area of research.

\vspace{3pt}
\noindent\textbf{Bivariate Bicycle Codes.} The recent seminal work~\cite{bravyi2024high} introduces a family of qLDPC codes called \emph{bivariate bicycle} (BB) codes, which possess desirable properties including high decoding rates, large code distances, and relatively simple structures, making them promising candidates for near-term qLDPC QEC architectures. We provide an overview of its structure.

\begin{figure}[ht]
\centering
\includegraphics[scale=0.2]{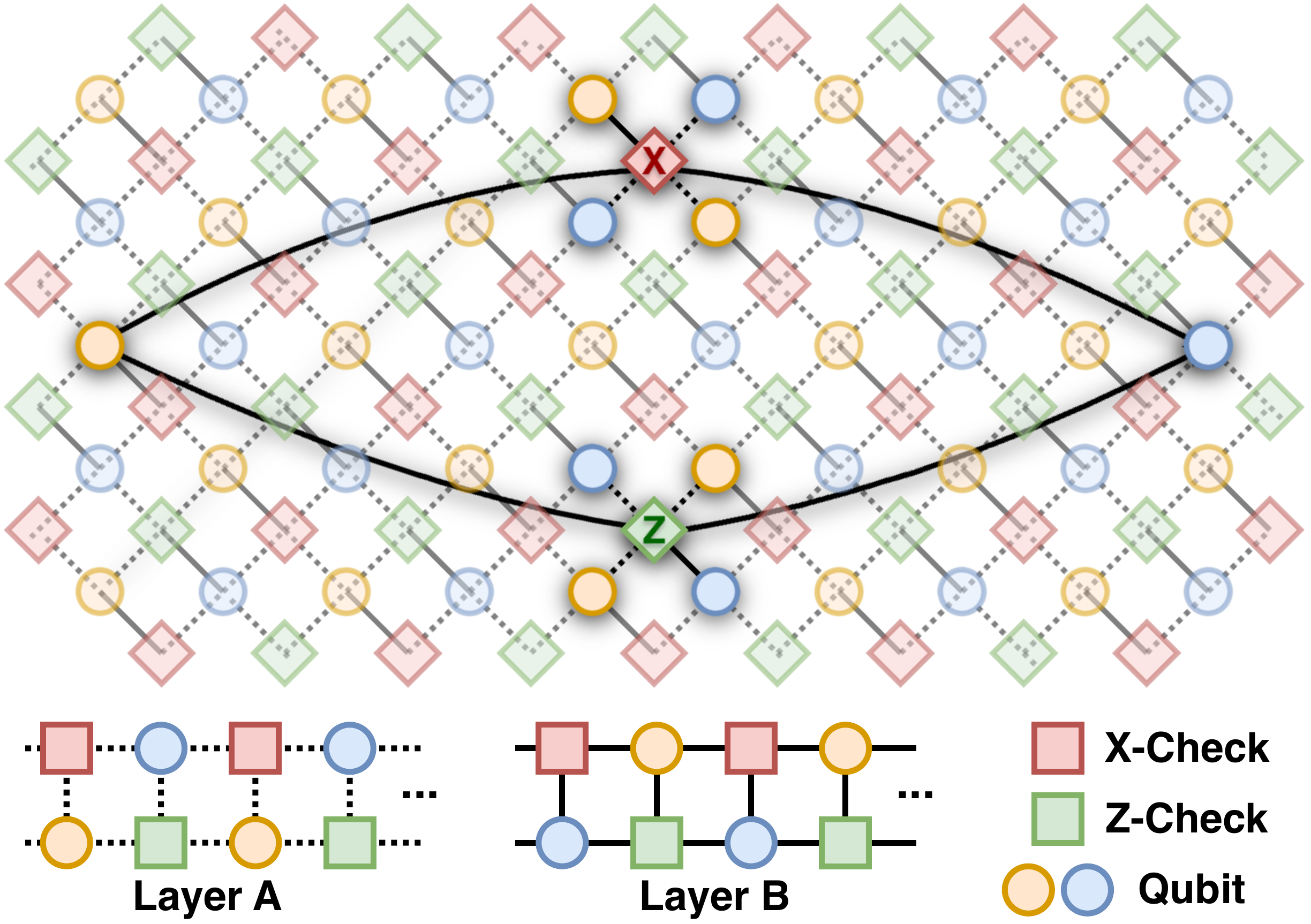}
\caption{Structure of BB code.}
\label{fig: BB}
\vspace{-8pt}
\end{figure}

Fig.~\ref{fig: BB} shows the Tanner graph of a BB code instance, featuring two types of parity checks: $X$- and $Z$-checks, composed solely of $X$ and $Z$ Pauli operators, respectively. Each $X$- or $Z$-check node connects to six variable nodes, classified into two layers, A and B, indicated by dashed and solid lines in Fig.~\ref{fig: BB}. This layering creates four adjacent edges and two non-local edges. Additionally, each qubit participates in three $X$-checks and three $Z$-checks—a higher degree that prevents the use of MWPM decoders~\cite{higgott2022pymatching}, which require each qubit to have a degree of two. The current decoder used for BB codes, BP+OSD (introduced in Sec.~\ref{subsec: BP+OSD}), is unsuitable for real-time decoding due to its inherently sequential nature and the high computational cost of matrix inversion, as detailed in Sec.~\ref{subsec: BP+OSD}.


\begin{figure*}[!ht]
    \centering
    \includegraphics[width=1\textwidth]{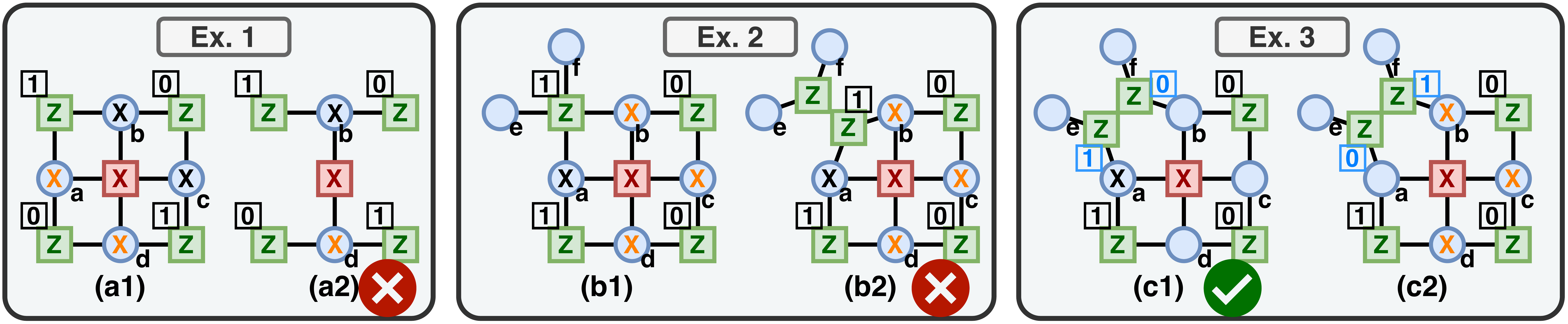}
    \caption{(a) Deleting nodes and edges may cause failure: no error patterns on the remaining qubits produce the observed syndromes in (a2). (b) Not all syndrome splits break quantum degeneracy: the syndrome split in (b2) is ineffective because two distinct error patterns (black and orange) still yield the same error syndromes. (c) Different syndrome assignments can guide BP to converge to distinct error patterns: ideally, the new syndromes lead to the minimal-weight error, as shown in (c1).}
    \label{fig:motivatingEx}
\end{figure*}

\subsection{BP+OSD Decoder}{\label{subsec: BP+OSD}}
This subsection presents the leading BP+OSD decoder for qLDPC codes along with an analysis on its limitations.

\vspace{3pt}
\noindent\textbf{Belief Propagation (BP).} BP is an effective decoding algorithm for classical LDPC codes~\cite{richardson2018design, mackay1997near}. It iteratively “passes” messages, represented by specific quantities, between neighboring nodes in the Tanner graph. After each iteration, the most recent messages are used to estimate the error probability (EP) for each variable node. An EP above 0.5 suggests an error; otherwise, it does not. For detailed EP updates during BP iterations, see~\cite{roffe2020decoding, higgott2024practical}. BP is efficient, with $O(n)$ complexity~\cite{mackay1997near}, supports parallelization, and has efficient hardware implementation~\cite{valls2021syndrome}.

However, applying BP to qLDPC codes presents challenges. Ideally, EPs should converge to 0 or 1, providing a clear error estimate. Yet, quantum degeneracy in qLDPC codes results in multiple error patterns with equal probability, where BP fail to distinguish a single solution. Consequently, EPs may stagnate around 0.5 or fail to converge accurately~\cite{roffe2020decoding}, leading to unreliable error estimates and decoding failures. Post-processing techniques, such as order statistics decoding (OSD)~\cite{roffe2020decoding, panteleev2021degenerate}, are often needed to address BP’s limitations and is the most accurate approach currently available.

\vspace{3pt}
\noindent\textbf{Order Statistics Decoding (OSD).} OSD is employed when BP fails to produce error estimations consistent with the observed syndromes. It retains the most reliable estimations—those with EPs close to 0 or 1—and revises the remaining estimates to align with the error syndromes. Since searching all possible revisions incurs exponential overhead, solutions like \emph{OSD-0}~\cite{roffe2020decoding} limit the set of qubits considered for revision. They incrementally expand the set of qubits $\mathcal{J}$ to be revised in ascending order of their reliability, until a set is found that, when revised, matches the observed syndromes. Variants such as \emph{OSD-CS}~\cite{roffe2020decoding} considers larger qubit sets, enabling the correction of more complex error patterns but at the cost of increased computation time. These two OSD-based methods strike a balance between computational complexity and accuracy, making them the mainstream solutions for qLDPC decoding.

However, prior studies show that these methods are still impractical for real-time implementation, even for codes of modest size (e.g., $\sim 100$ qubits)~\cite{valls2021syndrome}, let alone the larger codes required for practical quantum computing. There are two primary reasons: (1) The incremental expansion of $\mathcal{J}$ is inherently \emph{sequential}, severely limiting parallelization opportunities~\cite{valls2021syndrome}. (2) Verifying whether a given $\mathcal{J}$ yields a valid revised error estimation requires computationally \emph{expensive matrix inversion~}\cite{iolius2023decoding}. Developing decoders that bypass these computational bottlenecks while preserving high accuracy is essential for the feasibility of qLDPC codes~\cite{valls2021syndrome}.

\section{Motivation}

In this section, we present several examples showing the complexities of resolving the issue of quantum degeneracy, which finally motivates our solution.

\vspace{3pt}
\noindent\textbf{Example 1: Deleting Nodes and Edges May Cause Failure.} Fig.\ref{fig:motivatingEx}(a1) illustrates an example of quantum degeneracy, where two error patterns (black and orange) produce the same error syndromes. Deleting certain variable nodes along with their connected edges can break the degeneracy. As shown in Fig.\ref{fig:motivatingEx}(a2), removing nodes $a$ and $c$ results in the remaining portions of the two error patterns generating different syndromes. This occurs because the black (resp. orange) $X$-error on qubit $b$ (or $d$) flips only the two upper (or lower) $Z$-checks. However, no error patterns on the remaining qubits $b$ and $d$ can match the original syndrome.

\vspace{3pt}
\noindent\textbf{Takeaway.} Deleting nodes and edges may not be an effective method for breaking degeneracy, because the remaining graph may not support an error pattern that match the observed syndromes. This motivates us to explore syndrome splitting that preserves the graph structure to the best extent.

\vspace{3pt}
\noindent\textbf{Example 2: Not All Syndrome Splits Break Degeneracy.} In Fig.\ref{fig:motivatingEx}(b1), two error patterns (black and orange) produce the same error syndromes. To break this degeneracy, we attempt to split the upper-left syndrome. However, an arbitrary syndrome split may not be sufficient to break the degeneracy. For instance, the split in Fig.\ref{fig:motivatingEx}(b2) fails to do so, as the two error patterns still yield identical syndromes after the split.

\vspace{3pt}
\noindent\textbf{Takeaway.} The syndrome split operation must be carefully designed to successfully break degeneracy.

\vspace{3pt}
\noindent\textbf{Example 3: Different Syndrome Assignments Lead to Distinct Error Patterns.} Suppose we have split a syndrome node into two new ones, successfully breaking the degeneracy. Fig.\ref{fig:motivatingEx}(c1)(c2) provides an example, where the two error patterns (black and orange) now produce distinct syndromes. To complete the syndrome split, however, we must assign appropriate values (0 or 1) to these two new nodes. This choice of assignment will guide BP to converge to the corresponding error pattern. For instance, if we assign new syndromes as in Fig.\ref{fig:motivatingEx}(c1) (resp. (c2)), the BP in subsequent rounds will converge on the black (resp. orange) error pattern. Our goal is to select the syndrome assignment that most likely captures the actual underlying errors.

\vspace{3pt}
\noindent\textbf{Takeaway.} The syndromes assigned to split nodes must be carefully chosen to approximate the true error pattern.

These three examples illustrate the core idea behind our solution: to design a syndrome split operation that effectively removes degeneracy while providing reliable estimates for the new syndromes, guiding subsequent BP rounds toward a desired error estimation. 

\section{Breaking Quantum Degeneracy}{\label{sec: break dengen}}
This section presents how to design an appropriate syndrome split that can resolve the issue of quantum degeneracy. This syndrome split technique is the core of our decoder.

\subsection{Characterization of Quantum Degeneracy}
This subsection rigorously formulate the issue of quantum degeneracy, which sets a foundation for developing syndrome splitting operations to break this degeneracy.

We first introduce \emph{parity matrices}, which efficiently represent parity checks in linear algebraic terms. We focus on CSS codes~\cite{calderbank1996good, steane1996multiple}, a general QEC code class that encompasses most of the qLDPC codes of interest. They include two types of parity checks: $X$- and $Z$-checks, consisting solely of $X$ and $Z$ Pauli operators, respectively. These checks are represented as row vectors of length $n$, where entries of $1$ indicate the qubits involved and $0$ indicate those not involved. By concatenating these vectors, we form the parity check matrices $H_X$ and $H_Z$, which specify all $X$- and $Z$-checks, respectively. Since all stabilizers must commute, the rows of $H_X$ are orthogonal to all rows of $H_Z$. This orthogonality condition can be compactly expressed as $H_Z H_X^{T} = 0$. We decode $X$-errors and $Z$-errors separately on the $Z$-decoding graph and $X$-decoding graph, which correspond to the parity check matrices $H_Z$ and $H_X$, respectively. Due to the symmetry between $X$- and $Z$-decoding graphs, we focus on $X$-errors on the $Z$-decoding graph without loss of generality. An $X$-type error is represented by an $n$-dimensional binary vector $e$, where the $i$-th element being $1$ (resp. $0$) indicates the presence (resp. absence) of an $X$-error on the $i$-th qubit. The corresponding error syndrome is given by $s_Z = H_Z e$, where the $k$-th element of $s_Z$ represents the syndrome on the $k$-th parity check.

Quantum degeneracy occurs when two distinct error vectors $e_1 \neq e_2$ produce the same error syndrome: $H_Z e_1 = H_Z e_2 \Leftrightarrow H_Z (e_1 \oplus e_2) = 0$. This implies that $e_1 \oplus e_2$ is orthogonal to each row vector of $H_Z$. Consequently, $e_1 \oplus e_2$ must either correspond to an $X$-logical error (which cannot be detected by $Z$-checks) or belong to the row space of the $X$-parity check matrix $H_X$. Assuming the latter, as logical errors are not correctable, there exists an $X$-check $g_X \in \text{row}(H_X)$  such that $e_1 \oplus e_2 = g_X$, which implies $e_1 = e_2 \oplus g_X$.

In summary, the root cause of degeneracy is when two distinct errors differ by an $X$-check. Fig.~\ref{fig:motivatingEx}(a1) illustrates this with an example where two errors, $e_1 = X_b X_c$ (black) and $e_2 = X_a X_d$ (orange), result in the same error syndrome. The degeneracy arises because they differ by an $X$-check  $g_X = X_a X_b X_c X_d$.


\subsection{Breaking Degeneracy via Syndrome Split}{\label{subsec: intuition for split}}
We aim to break degeneracy while preserving as much error information as possible. To achieve this, we split the syndrome appropriately without removing any qubits, checks, or edges. Removing these elements could result in overly radical modifications to the Tanner graph, making it impossible to find a desired error estimation, as shown in Fig.~\ref{fig:motivatingEx}(a1)(a2). However, as illustrated in Fig.~\ref{fig:motivatingEx}(b1)(b2), not all methods of splitting a syndrome successfully break degeneracy. The key to a successful syndrome split lies in creating new parity checks with syndromes that differentiate the error patterns that originally corresponded to the same syndromes. Fig.~\ref{fig:motivatingEx}(c1)(c2) provides two examples of effective syndrome splits. We formalize this concept mathematically, which informs the design of an effective syndrome split operation.

\begin{figure}[ht]
\centering
\includegraphics[scale=0.18]{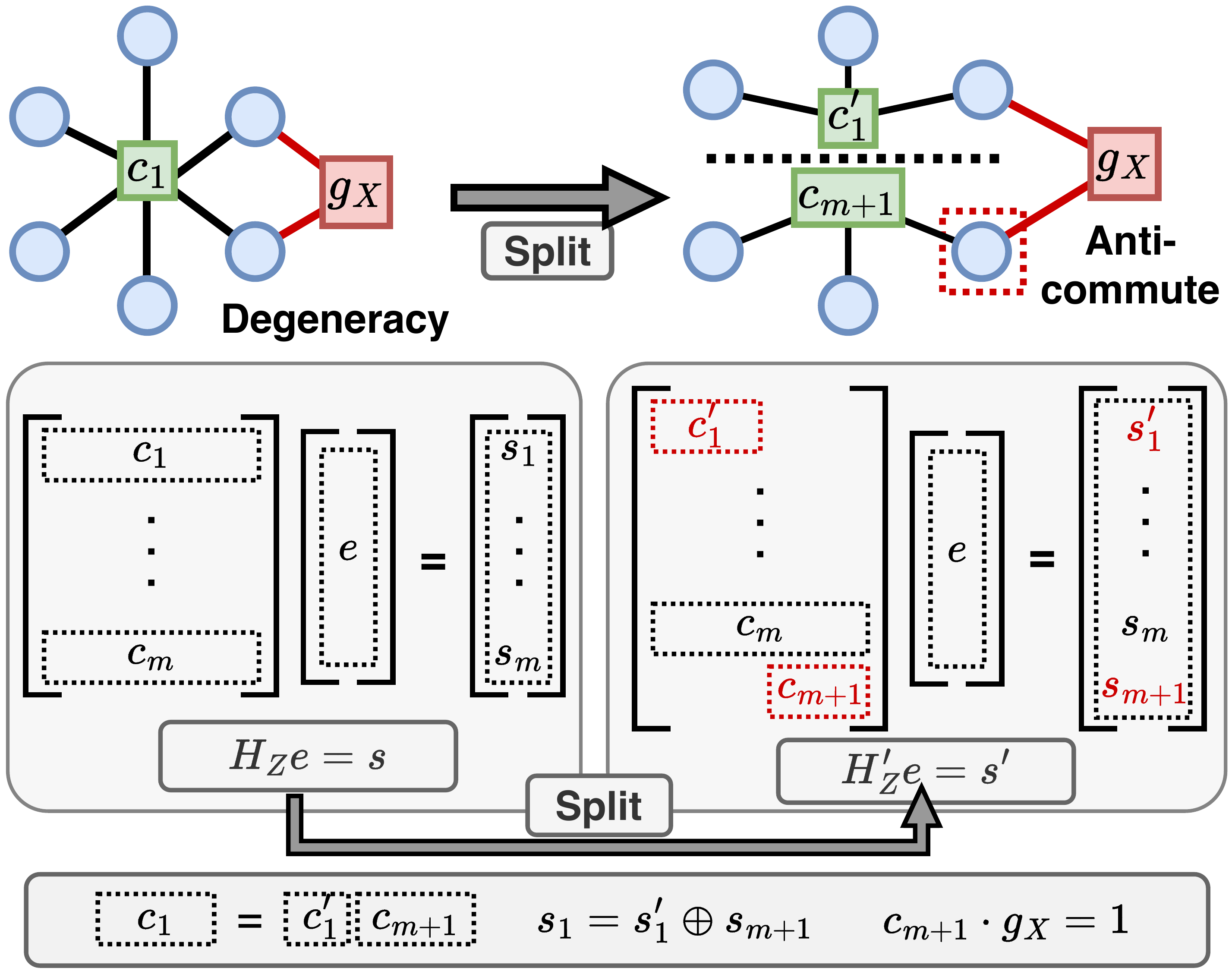}
\caption{Mathematical formulation of a desired syndrome split that breaks degeneracy. The check $c_1$ is split into $c_1^{\prime}$ and $c_{m+1}$ and we require $c_{m+1}$ to anti-commute with the $X$-check $g_X$ to break the associated degeneracy.}
\label{fig: breakDegeneracy}
\vspace{-8pt}
\end{figure}

We aim to break the symmetry caused by $g_X$ and distinguish between the two error patterns $e_1$ and $e_2$, where $e_1 = e_2 \oplus g_X$. As shown in Fig.~\ref{fig: breakDegeneracy}, the syndrome split operation can be effectively viewed as adding an additional constraint to the original linear equation  $H_Z e = s$. The original check $c_1$ is split into two checks, $c_1^{\prime}$ and $c_{m+1}$, and the original syndrome $s_1$ is split into $s_1^{\prime}$ and $s_{m+1}$, corresponding to $c_1^{\prime}$ and $c_{m+1}$, respectively. This is equivalent as adding a new equation regarding $c_{m+1}$ to the linear system and subtract that row to the one regarding $c_1$, which gives $c_1^{\prime}$. Consistency with this row operation naturally requires $s_1 = s_1^{\prime} \oplus s_{m+1}$.

To ensure that this syndrome split effectively breaks the degeneracy caused by $g_X$, the newly added $Z$-check $c_{m+1}$ must \emph{anti-commute} with $g_X$, satisfying $c_{m+1} \cdot g_X = 1$. This ensures that $e_1$ and $e_2$ can be distinguished by the syndromes on $c_1$ and $c_{m+1}$:
\begin{equation*}
\begin{cases}
c_1 \cdot e_1 = s_1, \\
c_{m+1} \cdot e_1 = s_{m+1},
\end{cases}
\begin{cases}
c_1 \cdot e_2 = s_1, \\
c_{m+1} \cdot e_2 = c_{m+1} \cdot (e_1 \oplus g_X) = s_{m+1} \oplus 1.
\end{cases}
\end{equation*}
In other words, while $e_1$ and $e_2$ are indistinguishable using the original $m$ parity checks, the newly added $(m+1)$-th check distinguishes them by assigning different syndromes. Assigning $c_{m+1}$ the syndrome $s_{m+1}$ guides BP to converge to $e_1$, while assigning it $s_{m+1} \oplus 1$ guides BP to converge to $e_2$. Thus, this syndrome split operation resolves the issue of quantum degeneracy effectively.




\subsection{Syndrome Split Operation}{\label{subsec: syn split operation}}
We have derived the condition for a syndrome split to break degeneracy: the newly added $Z$-check from the split must anti-commute with the $X$-check causing the degeneracy. Building on this insight, this subsection details our syndrome split operation, which determines two things: (1) \emph{Qubit distribution}, where the qubits connected to the original check are redistributed to the two newly created checks, and (2) \emph{Syndrome assignment}, which assigns appropriate syndromes to these new checks. Both components are designed depending on a metric called \emph{reliability}, which we introduce next.

\vspace{3pt}
\noindent\textbf{Reliability Function.} In state-of-the-art practical BP algorithms, each qubit $q$ is assigned a log-likelihood ratio (LLR) $L(q)$, which reflects the estimated error probability $p_q$ through the formula $L(q) = \log \left(\frac{1-p_q}{p_q}\right)$. A large positive LLR indicates a low error probability (EP), suggesting that the qubit is unlikely to support an error. Conversely, a large negative LLR implies a high EP, indicating the qubit is likely to support an error. The \emph{sign} of the LLR determines the \emph{direction} of the error estimation (whether the qubit supports an error or not), while the \emph{magnitude} represents the \emph{reliability} of this estimation. An LLR of zero corresponds to an EP of 0.5, indicating no reliability in the error estimation. Consequently, the reliability function for a qubit, $R(q)$, is defined as $R(q) = |L(q)|$, as provided by the current BP iteration. These reliability functions serve as key metrics for guiding qubit distribution and syndrome assignments, which we detail next.

\vspace{3pt}
\noindent\textbf{Syndrome Split Operation.} For a given $Z$-check $c$, we assume there exists an $X$-check $g_X$ overlapping with $c$, and aim to break the degeneracy caused by $g_X$ by splitting $c$ into two new checks, $c_1$ and $c_2$. Since $c$ must commute with $g_X$ by the construction of QEC codes, they must share an even number of overlapping qubits, denoted as $Q = \{q_1, \cdots, q_{2k}\}$. To ensure the split checks break degeneracy, $Q$ is distributed between $c_1$ and $c_2$ such that each new check involves an odd number of qubits from $Q$. This guarantees that $c_1$ and $c_2$ anti-commute with $g_X$, satisfying the condition for an effective syndrome split operation, as described in Sec.~\ref{subsec: intuition for split}. Specifically, if $k$ is odd, we let the two new checks each contain $k$ qubits. If $k$ is even, one check should contain $k-1$ qubits and the other $k+1$. This near-half split preserves the symmetry of the original and minimizes the disruption to its overall structure, potentially facilitating BP convergence in subsequent iterations. For example, in Fig.~\ref{fig: breakDegeneracy}, $k = 1$ is odd so we assign $k =1$ qubit to each new check $c_1^{\prime}$ and $c_{m+1}$.

Having determined the number of overlapping qubits for each new check, we propose two methods to explicitly assign the qubit distribution: (1) \emph{BP-guided}, and (2) \emph{syndrome-guided} approach.

\begin{figure*}[!ht]
    \centering
    \includegraphics[width=1\textwidth]{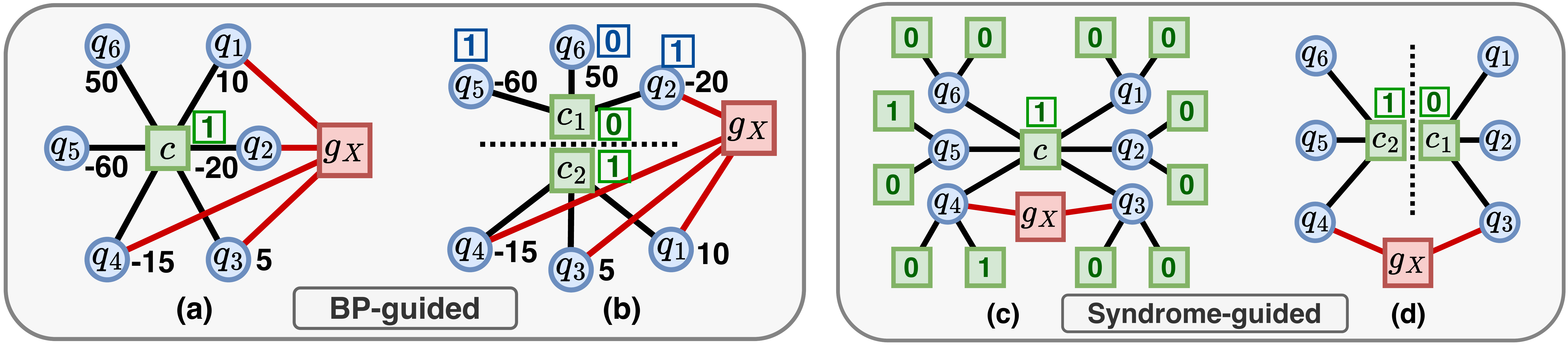}
    \caption{Two approaches for syndrome split operation. (a)(b) BP-guided approach. Assign the qubits with highest reliability to one check and determine its syndrome based on their LLRs. The syndrome of the other split check is decided accordingly. (c)(d) Syndrome-guided approach. Ensure that one side ($c_1$'s side in this case) has all-zero syndromes on the closest checks, and assign syndrome $0$ to that side.}
    \label{fig:SplitEx}
\end{figure*}

\vspace{3pt}
\noindent\textbf{(1) BP-Guided Approach.} This approach uses the reliability function to guide the distribution of qubits. The key idea is to assign qubits with high reliability to one of the new checks, ensuring higher confidence in determining its syndrome later. Following this principle, we select $k$ or $k-1$ (near-half) overlapping qubits with the highest reliability (depending on whether $k$ is odd or even) and assign them to the new check $c_1$. Next, we choose the non-overlapping qubits with the highest reliability from the original check $c$ and assign them to $c_1$ until the number of qubits involved in $c_1$ reaches half the total in $c$. For example, in Fig.~\ref{fig:SplitEx}(a), the $X$-check $g_X$ overlaps with four qubits ($q_1$-$q_4$) from the $Z$-check $c$ ($k = 2$), with their LLR values specified. Following the distribution method, we assign one overlapping qubit ($k-1 = 1$) with the highest reliability, $q_2$, to $c_1$. Then, two non-overlapping qubits with the highest reliability ($q_5$ and $q_6$) are also assigned to $c_1$. The remaining qubits ($q_1, q_3, q_4$) are assigned to $c_2$.

Next, we determine the syndromes of $c_1$ and $c_2$. Using the LLRs of qubits in $c_1$, we compute its syndrome because it includes qubits with the most reliable estimations. In Fig.~\ref{fig:SplitEx}(b), the LLRs $-60, 50, -20$ imply error estimates $1, 0, 1$, respectively, which sum to a syndrome of $0$ on $c_1$. To ensure the new syndromes of $c_1$ and $c_2$ sum to the original syndrome $1$, we assign $c_2$ the syndrome $1$.

\vspace{3pt}
\noindent\textbf{(2)  Syndrome-guided approach.} This approach determines the syndromes of $c_1$ and $c_2$ based on the syndromes of the closest checks connected to them. The key idea is that if the closest checks on one side, such as $c_1$’s side, all have syndrome $0$, it is likely that no error exists on the qubits connected to $c_1$, and $c_1$ should be assigned syndrome $0$. The syndrome for $c_2$ is then derived to ensure the new syndromes sum to the original syndrome. We stipulate that qubits are distributed evenly between $c_1$ and $c_2$ while satisfying the degeneracy-breaking condition. For instance, in Fig.~\ref{fig:SplitEx}(d), the qubit distribution satisfies this condition because the new check $c_1$ anti-commutes with $g_X$. Since the syndromes of the neighboring checks on $c_1$’s side are all $0$, we assign syndrome $0$ to $c_1$. We then assign syndrome $1$ to $c_2$.

\section{A Decoder Based on Syndrome Split}{\label{sec: check split}}
In this section, we present the \frameworkname~decoder, built upon the syndrome split operation introduced in Sec.~\ref{subsec: syn split operation}. In high level, we insert syndrome split operations into the standard BP iterations. Specifically, the \frameworkname~decoder operates iteratively through the following three steps: (1) \emph{BP iterations}, which calculate the LLR and error estimation for each qubit; (2) \emph{termination check}, which determines whether the current error estimation matches the observed syndrome; and (3) \emph{syndrome split}, which modifies the Tanner graph to improve BP’s convergence in subsequent iterations. Each step is detailed below along with a discussion on complexity. They are combined to give Algorithm~\ref{alg:jazzsplit_decoder}.

\vspace{3pt}

\noindent\textbf{Step 1. BP iterations. }
This step performs $m$ BP iterations to update the LLRs for each qubit, yielding error estimations and revealing their reliability. Qubits with unreliable estimations are more likely to be influenced by degeneracy issues. This information is crucial for identifying $X$-checks that may cause degeneracy and guides the subsequent syndrome split operation to resolve it. The parameter $m$ is chosen as a small number ($10\sim 20$) since it already reveals the error information that helps us to locate the degeneracy.  




\vspace{3pt}
\noindent\textbf{Complexity Discussion. }The complexity of each BP iteration is $O(n)$, and our approach maintains the same total number of iterations as standard BP. The difference lies in periodically inserting syndrome splits after every $m$ iterations. Consequently, the overall cost of BP iterations in our approach remains equivalent to that of standard BP. We clarify that while the syndrome split operation may increase the total number of nodes in the Tanner graph and increases $n$, this increase is limited to a constant factor of two, as each syndrome node is split into at most two new nodes.

\vspace{3pt}
\noindent\textbf{Step 2. Termination check. }
The syndrome split operation is intended to assist BP in producing increasingly accurate error estimations that progressively approach the target syndrome. Ideally, the difference between the given syndrome $s$ and the syndrome of the estimated error $\hat{s}$ should decrease with each iteration, eventually satisfying $s - \hat{s} = 0 \text{ mod } 2$. When this condition is met, decoding terminates as the error estimation matches the observed syndrome.

However, syndrome splitting may occasionally leave the number of mismatched syndromes, $d = |s - \hat{s}|$, unchanged or even increased. If this occurs repeatedly, it indicates that further syndrome splits are unlikely to improve results and will only prolong decoding time. To handle this, we introduce a monitoring quantity, $K$, initialized to zero. Each time $d$ remains constant or increases, $K$ is incremented by $1$. If $d$ decreases and $K > 0$, K is decremented by $1$. Decoding terminates when $K$ reaches a threshold of $3$, signaling that no further improvements are likely.

\vspace{3pt}
\noindent\textbf{Complexity Discussion. }
Two termination conditions are evaluated during each iteration. The first checks whether the current error estimation, $\hat{e}$, matches the observed syndrome $s$ by verifying $H_Z \hat{e} = s$. This involves matrix-vector multiplication with a complexity of $O(n)$, leveraging the sparsity of $H_Z$ and $e$. Specifically, each row of $H_Z$ has a bounded number of non-zero entries, and the total number of rows is at most $n$, as defined by qLDPC codes. The second condition monitors whether the quantity $K$, which tracks decoding stagnation, has reached a predefined threshold. This involves simple counter operations with negligible computational cost. Together, checking the termination conditions add ans $O(n)$ overhead to the decoding process.
\begin{algorithm}
\caption{\frameworkname~Decoder}
\label{alg:jazzsplit_decoder}
\KwData{Parity-check matrix $H_X$, $H_Z$, observed syndrome $s_Z$}
\KwResult{Estimated $X$-error pattern $\hat{e}$}
Initialize $K \gets 0$ \textit{\textbf{// Monitoring quantity for termination condition}}
Initialize Tanner graph based on $H_Z$, $H_X$\;
\While{true}{
    \textbf{Step 1: BP Iterations}\;
    Run BP iterations, update LLRs, and compute error estimations $\hat{e}$\;
    Compute the difference $d \gets |s_Z - H_Z\hat{e}|$\; \textit{\textbf{// Number of mismatched syndromes}}
    
    \textbf{Step 2: Check Termination Conditions}\;
    \If{$d = 0$}{
        \Return $\hat{e}$; \textit{\textbf{// Est. Error matches the syndrome}}\\
        \textbf{break}\;
    }
    \Else{
        Update $K$ depending on the change in $d$\;
    }
    \If{$K = \textit{K\_max}$}{
        \Return \texttt{Decoding Stop}\;
        \textbf{break};
    }
    
    \textbf{Step 3: Syndrome Split}\;
    Compute reliability $R(c)$ for all $X$-checks $c$\;
    Select $g_X \gets \arg\min R(c)$\; \textit{\textbf{// Select the least reliable $X$-check}}\\
    Choose a $Z$-check overlapping with $g_X$ and perform syndrome split operation\;
    Update Tanner graph with the split checks\;
}
\end{algorithm}

\vspace{0.3em}
\noindent\textbf{Step 3. Syndrome split. }
In this step, the $X$-check most likely causing degeneracy is identified and addressed through a targeted syndrome split operation. To quantify the likelihood of a check causing degeneracy, we define the \emph{reliability function for a check node} as the average reliability of its adjacent qubits:

$$R(c) = \frac{1}{C_q}\sum_{q\sim c} R(q)$$
where $q \sim c$ denotes the qubits connected to the check node $c$, and $C_q$ is the total number of these qubits. Intuitively, an $X$-check with low reliability is more likely to contribute to degeneracy (see Fig.~\ref{fig:intro}(b3) for an example).

We choose to address the degeneracy caused by $g_X$, the X-check with the lowest reliability. Specifically, a $Z$-check overlapping with $g_X$ is selected, and the syndrome split operation outlined in Sec.~\ref{subsec: syn split operation} is performed. The modified Tanner graph is then passed to subsequent BP iterations, encouraging better convergence.

\vspace{3pt}
\noindent\textbf{Complexity Discussion. }
This step begins by identifying the $X$-check with the lowest reliability. The reliability of each qubit is directly provided by the current BP iteration. Computing the reliability function for all $X$-checks can be done by $H_X r$, where $r$ is the reliability vector. This operation has complexity $O(n)$, similar to the $O(n)$ complexity of computing $H_Z\hat{e}$ in the termination check. Selecting the $X$-check with the lowest reliability involves scanning these values, which are bounded by $n$, adding at most $O(n)$. The actual syndrome split operation—whether guided by the BP-based or syndrome-based approach—relies on simple local calculations, including redistributing qubits between the split checks and assigning syndromes using readily available local information. Since qLDPC codes have bounded connectivity degrees, these operations incur a constant $O(1)$ overhead. Overall, the total complexity of this step is $O(n)$.

Combining the discussion of the three steps, we conclude that our decoder has a linear complexity. This claim will be further supported by the evaluation results presented in Sec.~\ref{sec: eval}.

\section{Code-Specific Decoder for BB Codes}
\label{sec: BB code decoder}
In this section, we introduce an alternative layered visualization of BB codes, distinct from the representation in the original paper~\cite{bravyi2024high}. This alternative visualization 
enhances the effectiveness of our decoder that combines BP and syndrome splitting (Sec.~\ref{subsec: bb code revisit}). We then present a decoder specifically tailored for BB codes.

\subsection{Revisiting BB Codes: An Alternative Perspective}{\label{subsec: bb code revisit}}
We introduce a new approach to interpreting the geometric structure of BB codes. In the original paper, qubits connected to each parity check (either an $X$- or $Z$-check) are divided into two layers, A and B, represented by dashed and solid lines (Fig.~\ref{fig: BB}). However, this classification results in qubits within each layer forming band shapes that consist of 4-cycles, which can degrade BP performance~\cite{higgott2024practical, panteleev2021degenerate}. Our alternative grouping organizes qubits into two layers, such that the Tanner graph within each layer consists of 6-cycles, allowing BP to yield more accurate results. Additionally, this classification provides a natural approach to syndrome splitting, leading to a simplified decoder tailored for BB codes.

\begin{figure}[ht]
\centering
\includegraphics[scale=0.17]{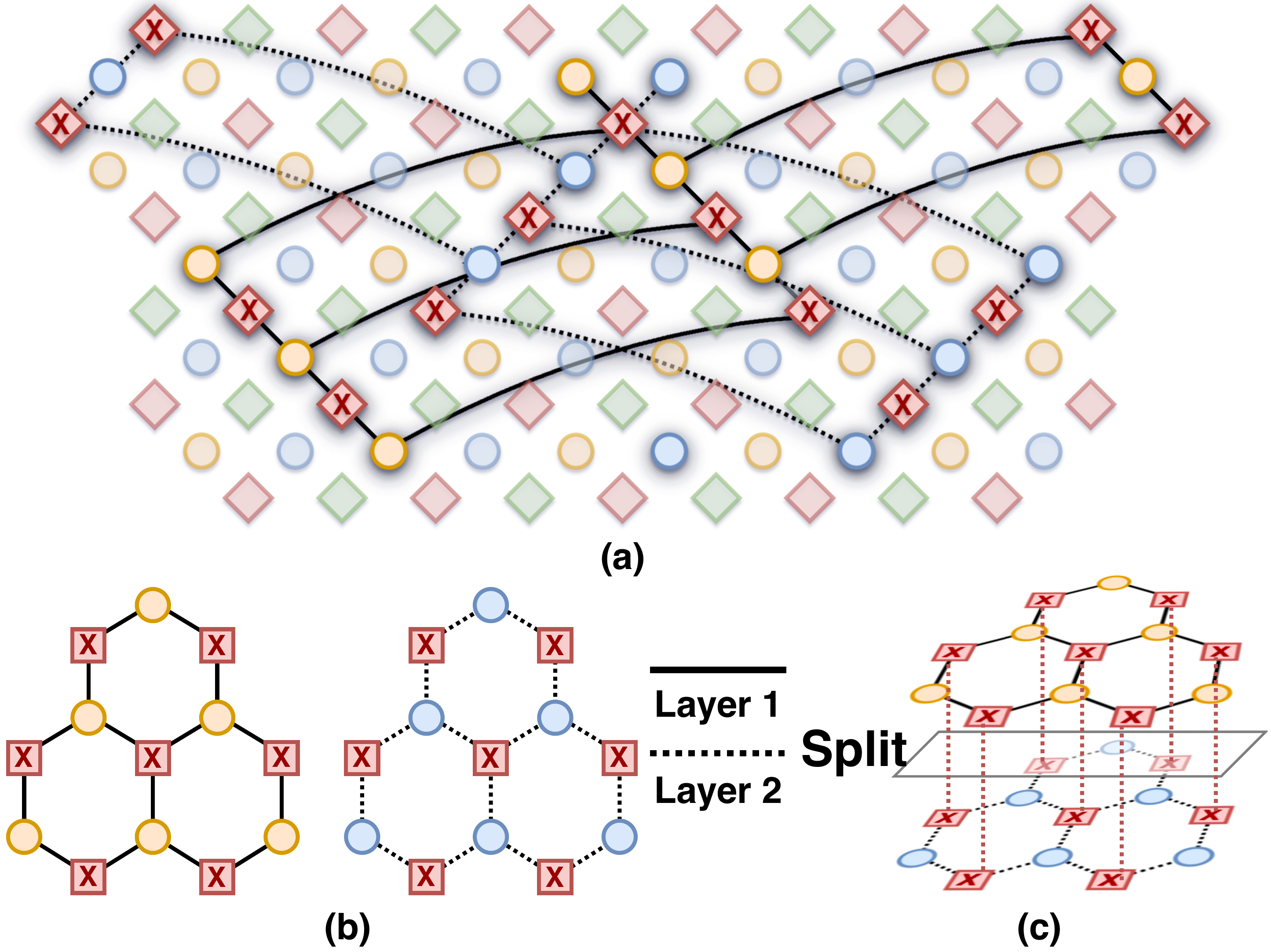}
\caption{(a) An alternative layering of BB code. The qubits connected to a check is classified into two layers specified by the solid and dashed lines. (2) This new layering gives a Tanner graph with 6-cycles instead of 4-cycles. (3) The layer structure leads to a straightforward syndrome split operation for the decoder.}
\label{fig: bbnew}
\vspace{-8pt}
\end{figure}

This new qubit grouping is shown in Fig.~\ref{fig: bbnew}(a). We regroup the yellow and blue qubits into two distinct layers. Extracting the qubits within each layer along with the edges between them, we obtain Tanner graphs that are equivalent to hexagonal tiling, as shown in Fig.~\ref{fig: bbnew}(b). Therefore, the shortest cycle length is 6, which is larger than in the original classification, thus enhancing BP performance~\cite{panteleev2021degenerate}. 
The figure only presents the case for $X$-checks, but the situation regarding $Z$-checks is completely symmetric.

\subsection{Decoder Tailored for BB Codes}
Based on the alternative qubit grouping into two layers, we design a straightforward decoder for BB codes. The syndrome-split operation becomes simplified: in the Tanner graph for $X$-checks, each $X$-check node is connected to six qubits—three yellow qubits in the upper layer and three blue qubits in the lower layer, as shown in Fig.~\ref{fig: bbnew}(c). When splitting the $X$-check, we assign the three yellow qubits to one of the new checks and the three blue qubits to the other. Syndromes are then assigned based on whether all adjacent check nodes on their respective sides have zero syndromes, as outlined in Sec.~\ref{subsec: syn split operation}. Importantly, this split method breaks the degeneracy caused by the connected $Z$-checks. Similarly, the same symmetric operation is applied for splitting checks in the Tanner graph of $Z$-checks. Despite its simplicity, this decoder 
enhances BP accuracy, achieving higher accuracy than the BP+OSD decoder despite reduced decoding time (Sec.\ref{sec:evaluation}).
\begin{figure*}[!tp]
        \centering
        \includegraphics[width=0.24\linewidth]{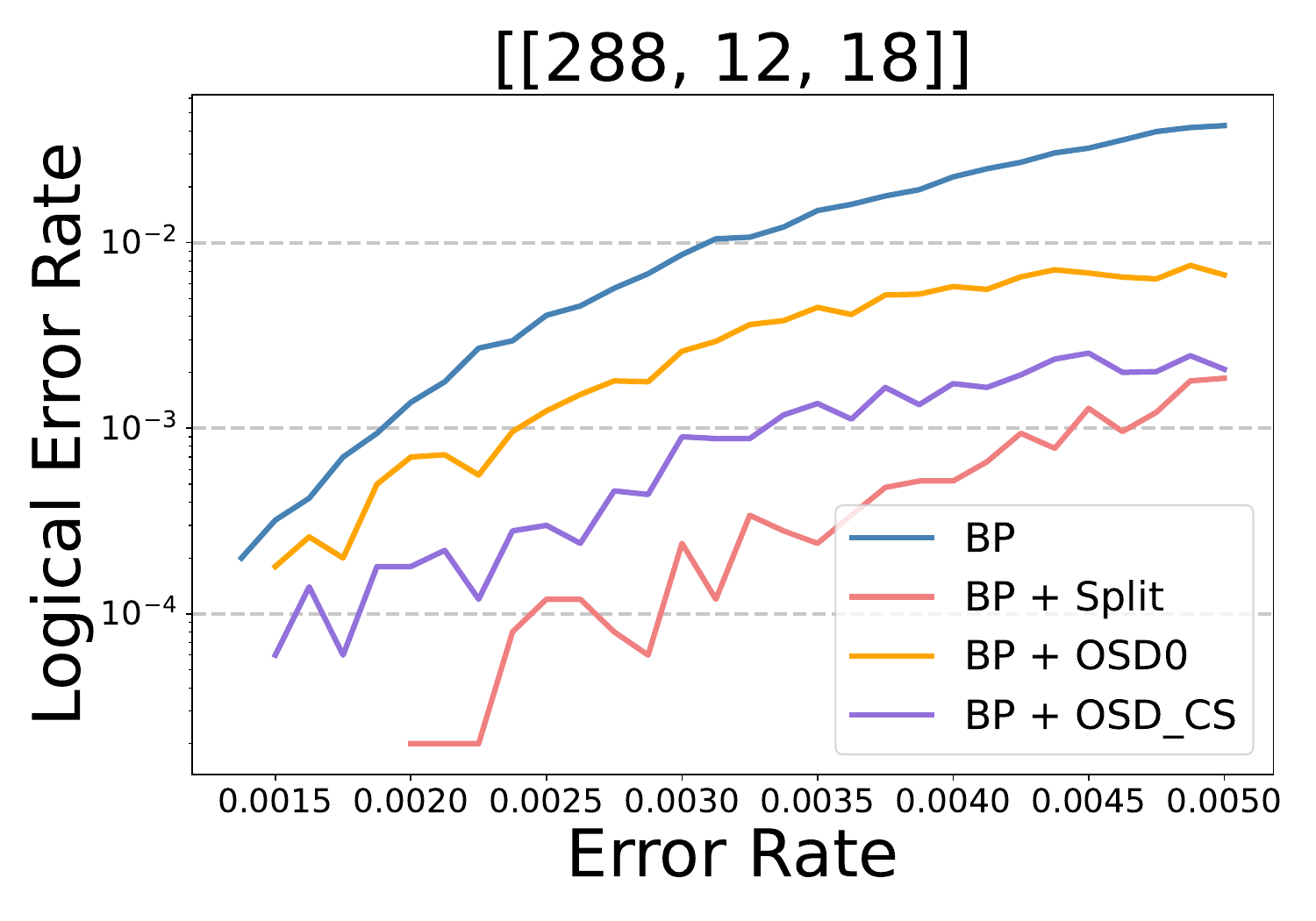} 
        \includegraphics[width=0.24\linewidth]{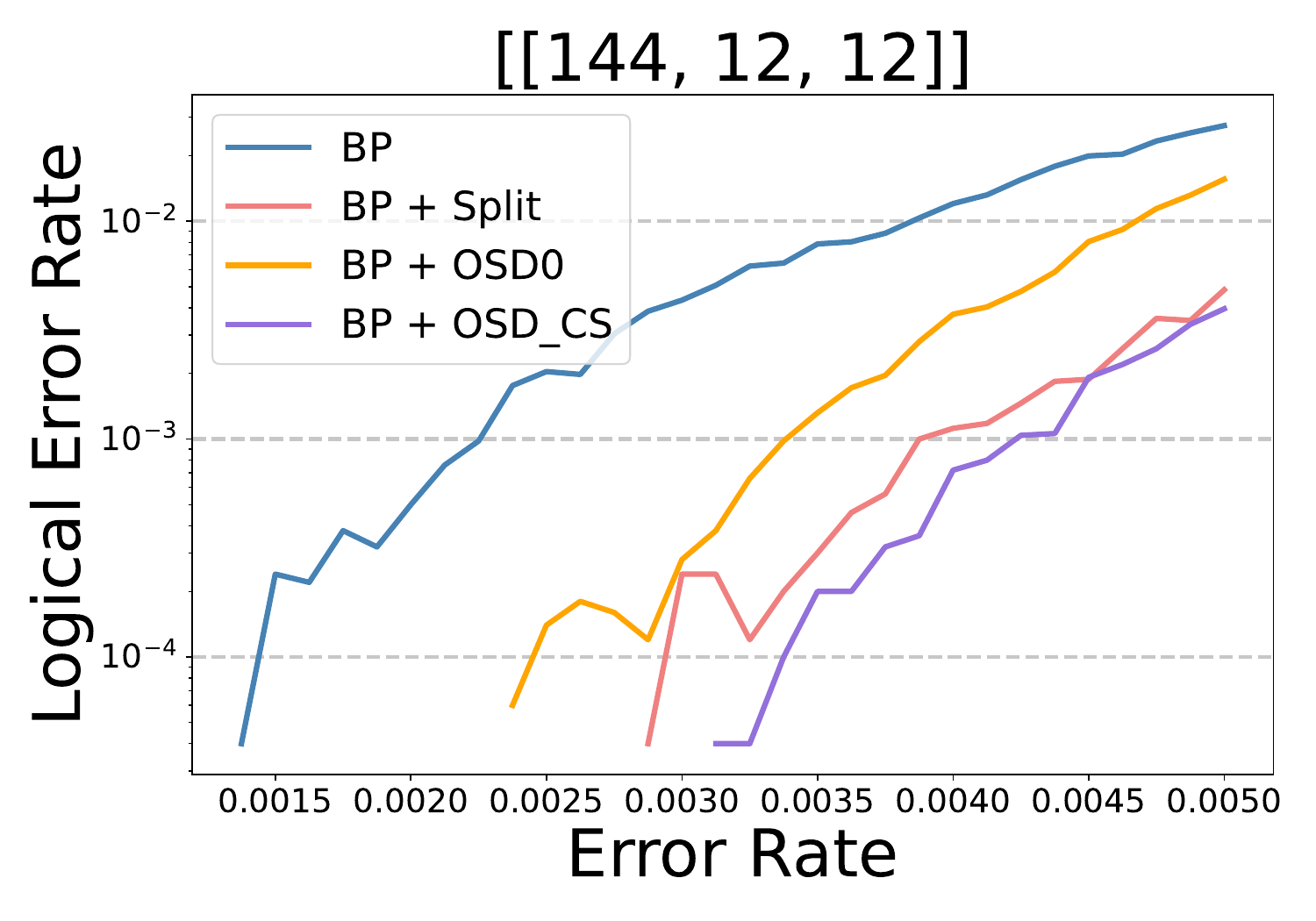}
        \includegraphics[width=0.24\linewidth]{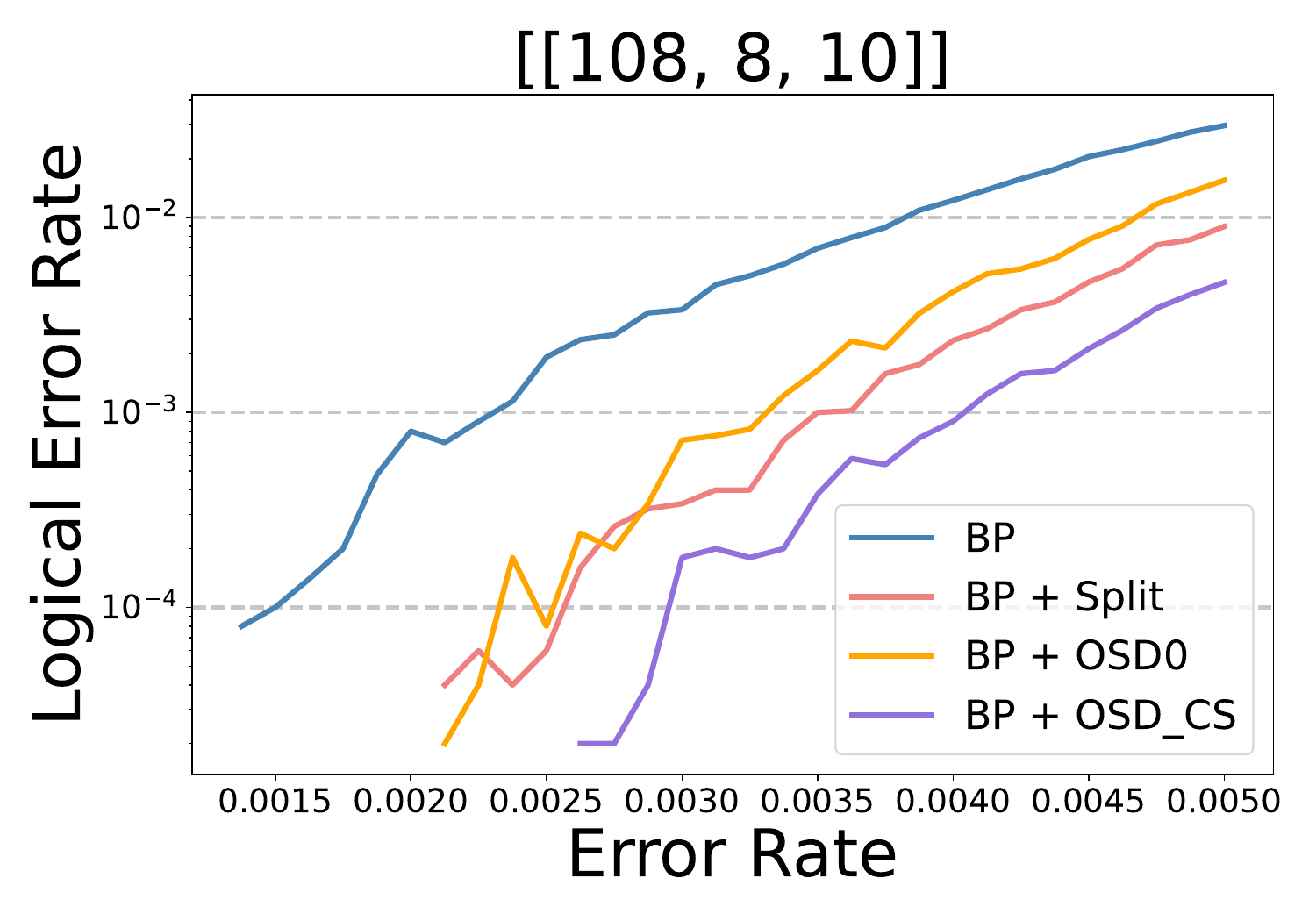}
        \includegraphics[width=0.24\linewidth]{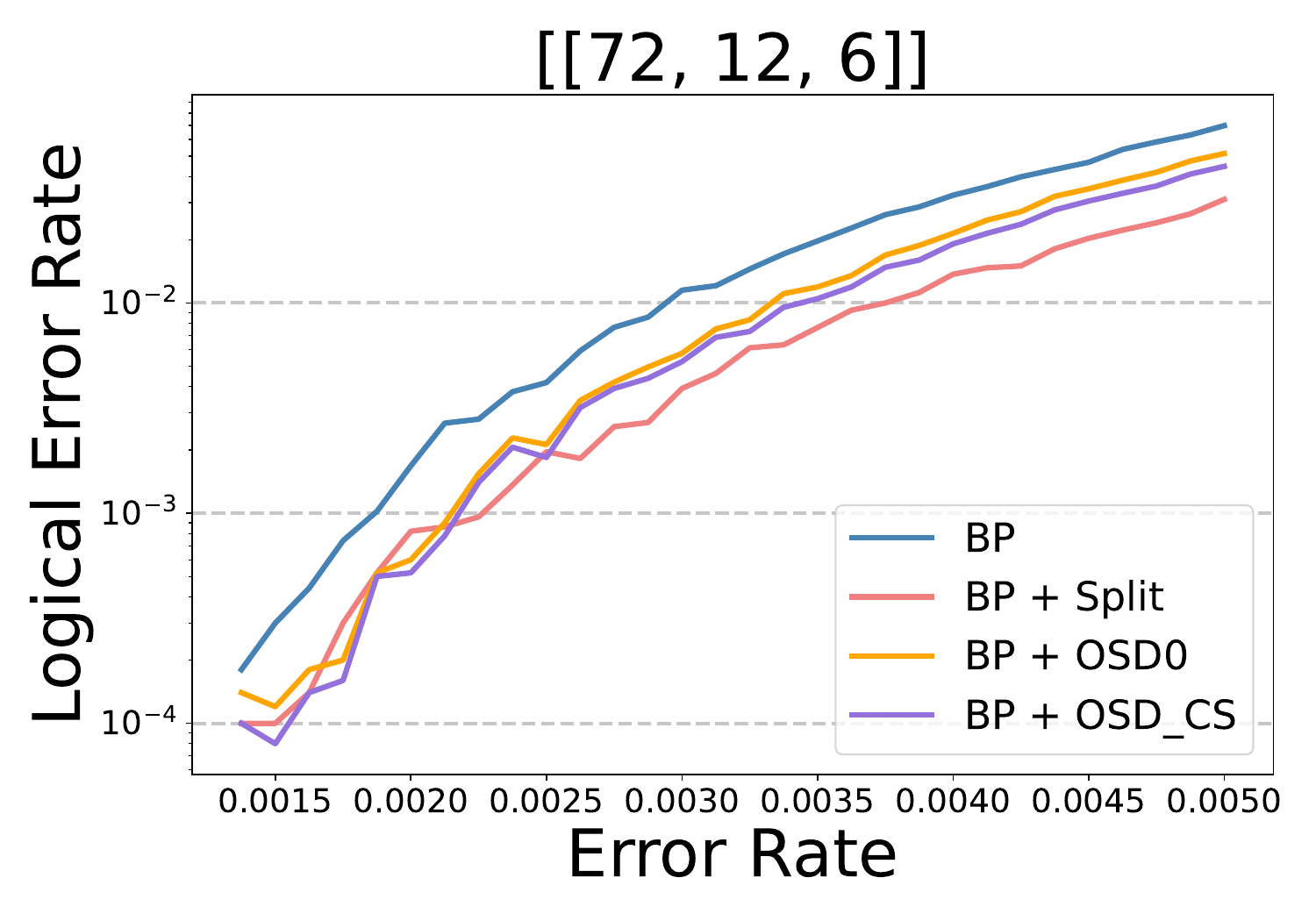}\\
        
        (a)\hspace{115pt}(b)\hspace{112pt}
        (c)\hspace{115pt}(d)

        \centering
        \includegraphics[width=0.24\linewidth]{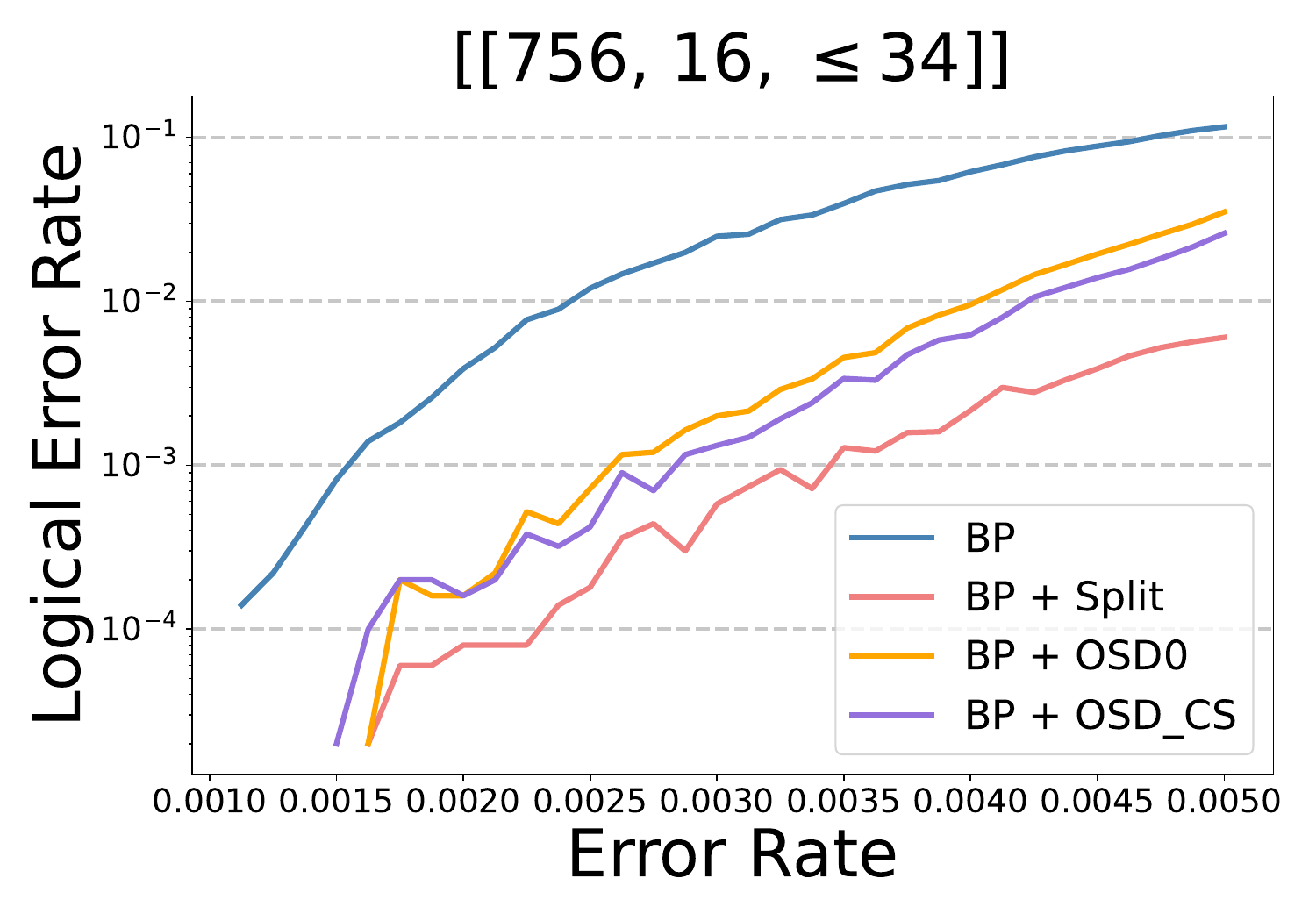} 
        \includegraphics[width=0.24\linewidth]{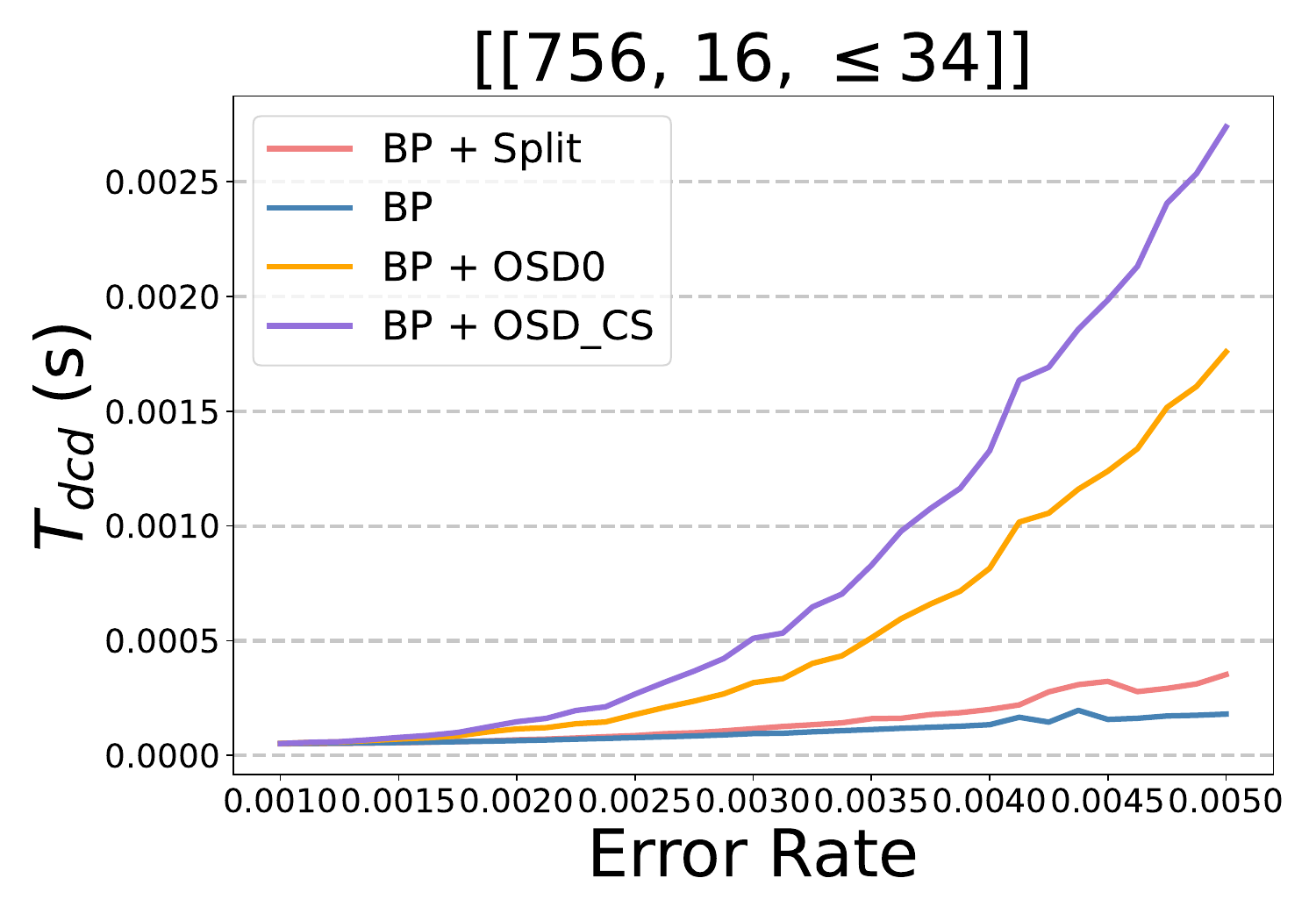}
        \includegraphics[width=0.24\linewidth]{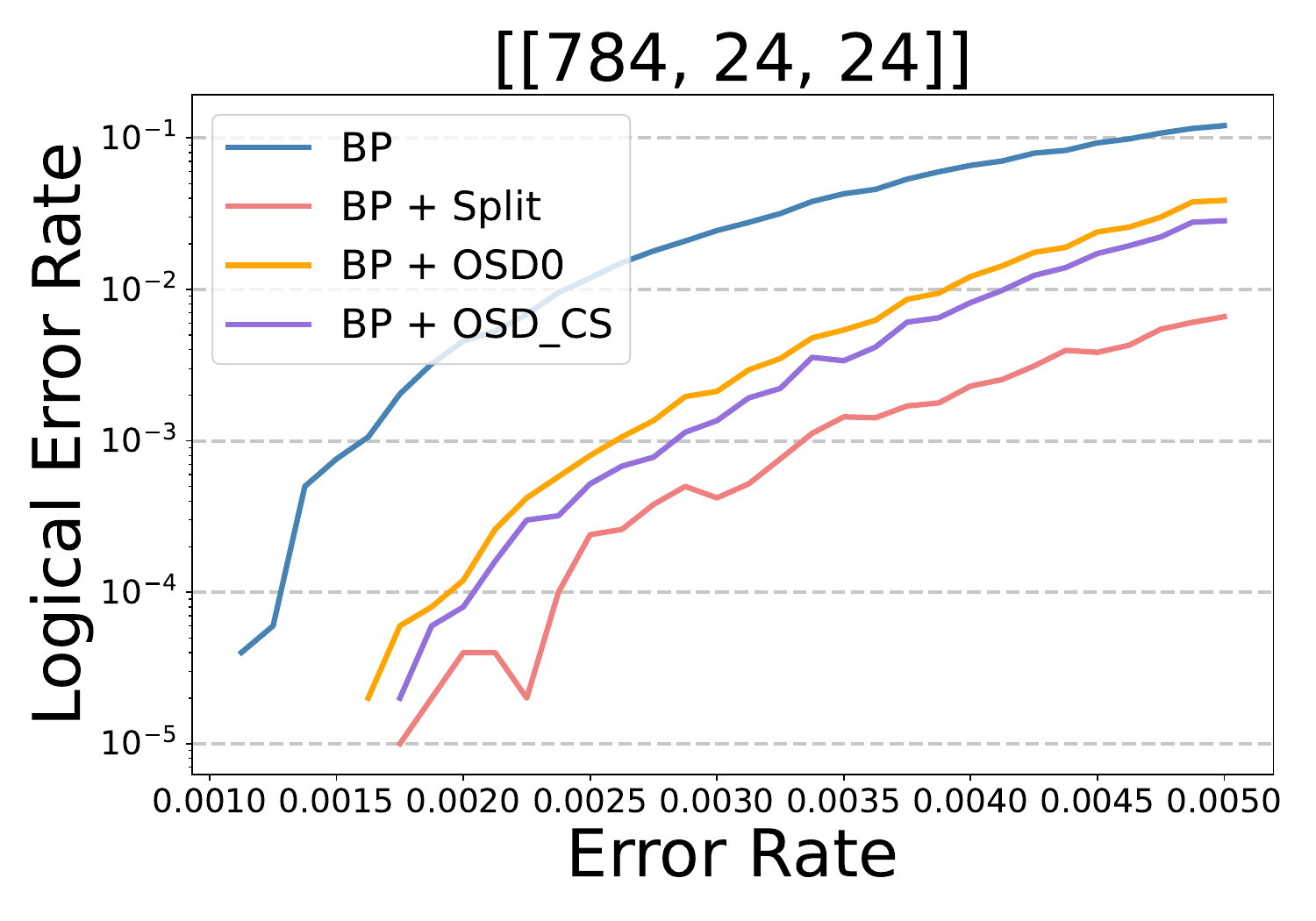}
        \includegraphics[width=0.24\linewidth]{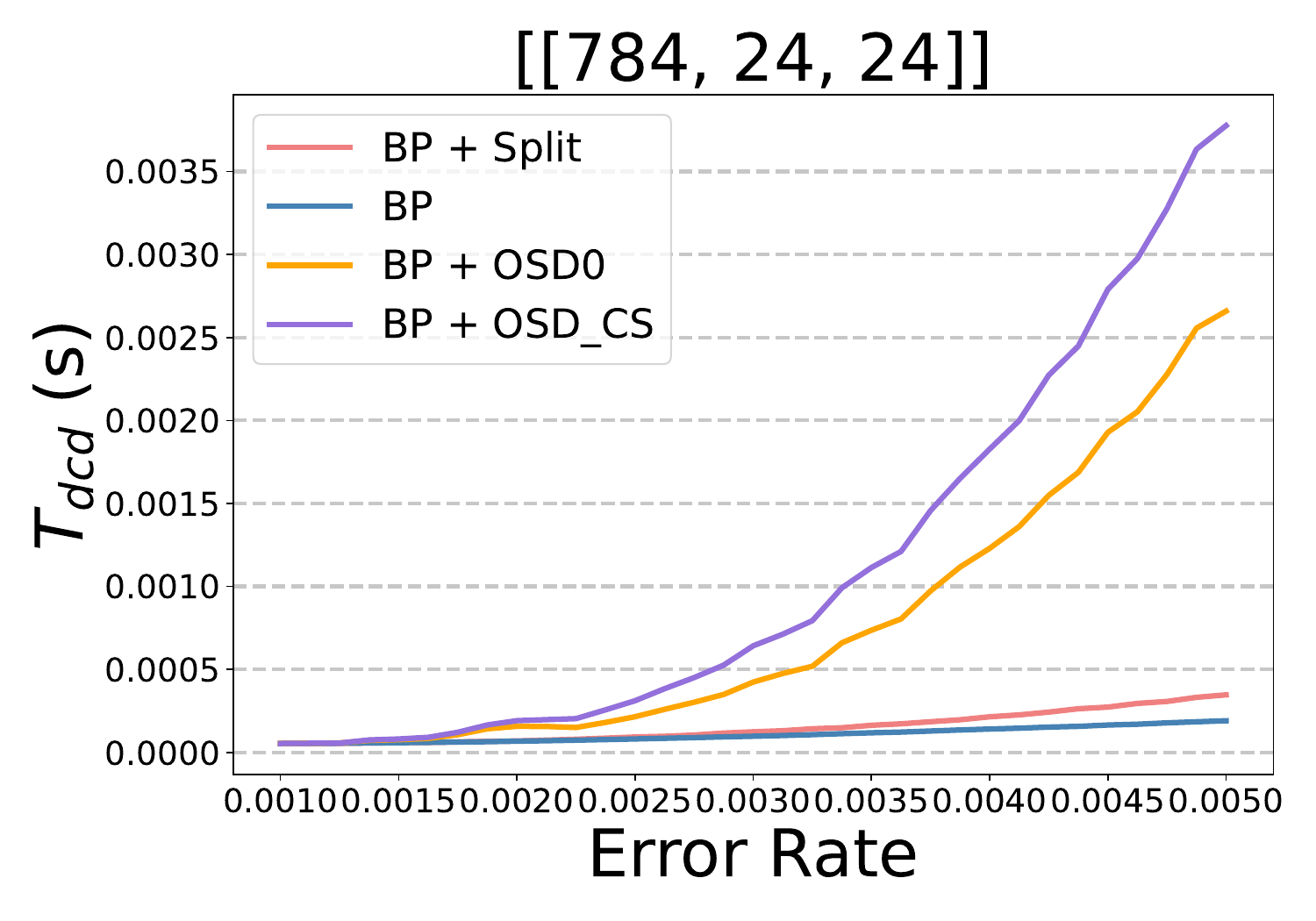}\\
        
        (e)\hspace{115pt}(f)\hspace{112pt}
        (g)\hspace{115pt}(h)
        \caption{\textbf{General Performance on BB Codes:} The notation \([[n, k, d]]\) represents a specific QEC code, where \(n\) denotes the number of data qubits, \(k\) indicates the number of encoded logical qubits, and \(d\) specifies the code distance. Figures (a)–(e) and (g) illustrate the performance of six BB codes with different  sizes, and Figures (f) and (h) present the decoding time ($T_{dcd}$) for two large-scale BB codes: [[756, 16, $\leq 34$]] and [[784, 24, 24]].}
        
       \label{fig:general_performance}
\end{figure*}

\section{Implementation and Evaluation}{\label{sec: eval}}
\label{sec:evaluation}

\begin{figure*}[!tp]
        \centering
        \includegraphics[width=0.32\linewidth]{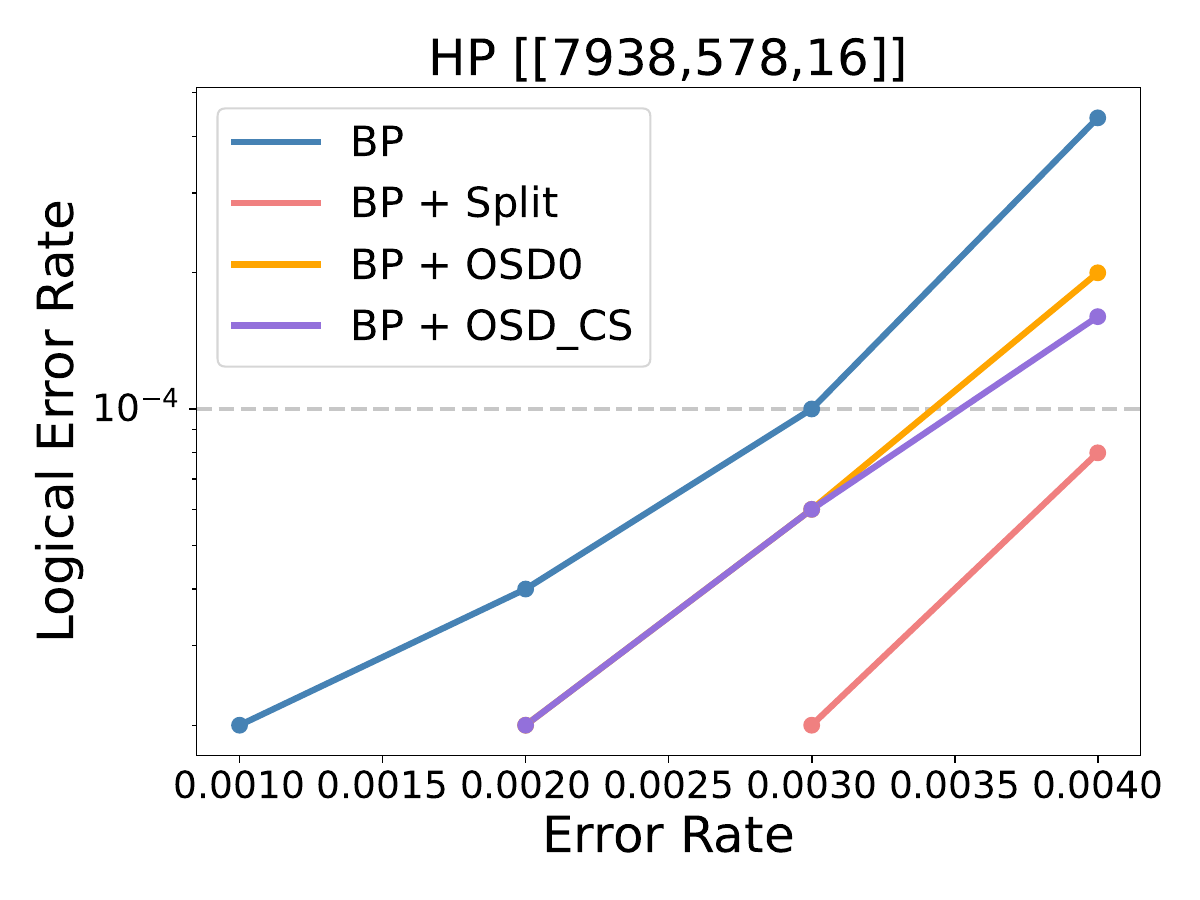}
        \includegraphics[width=0.32\linewidth]{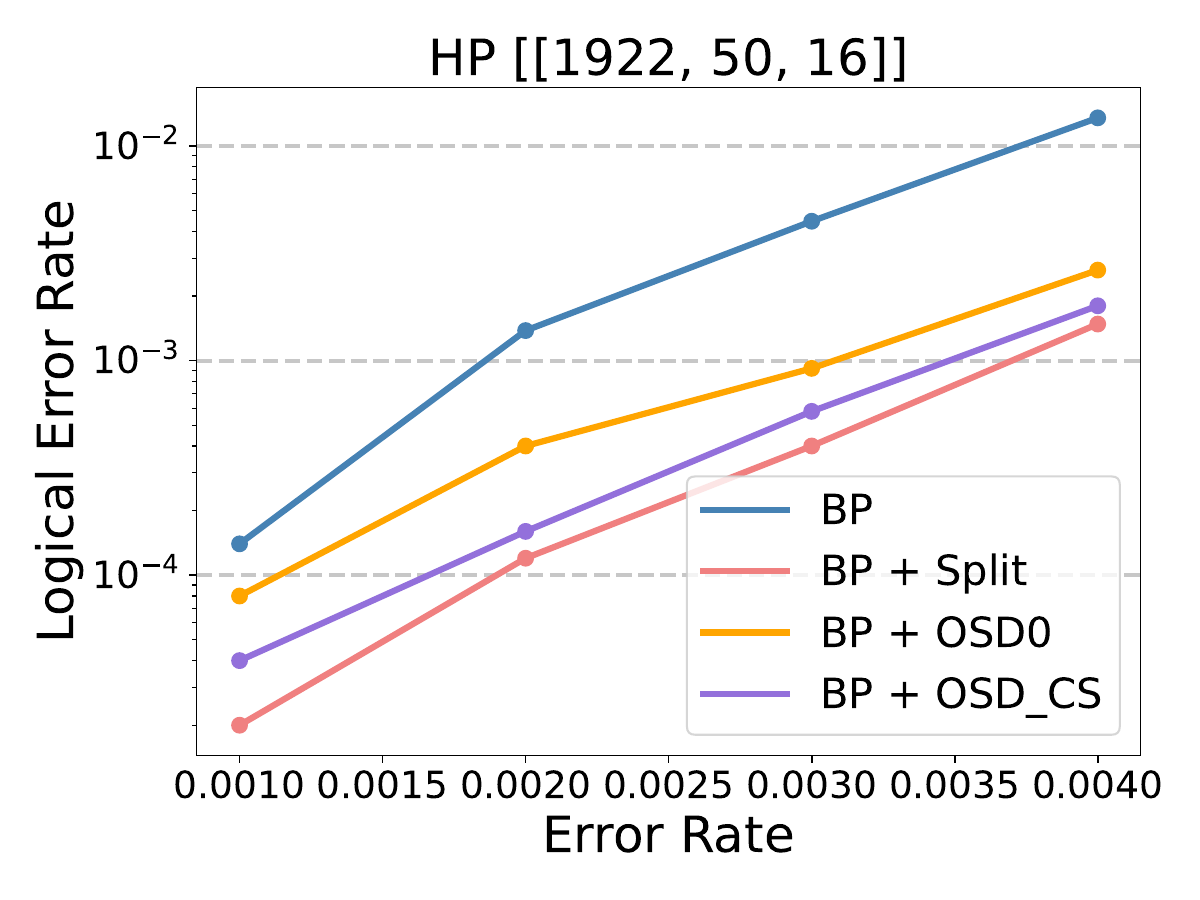}
        \includegraphics[width=0.32\linewidth]{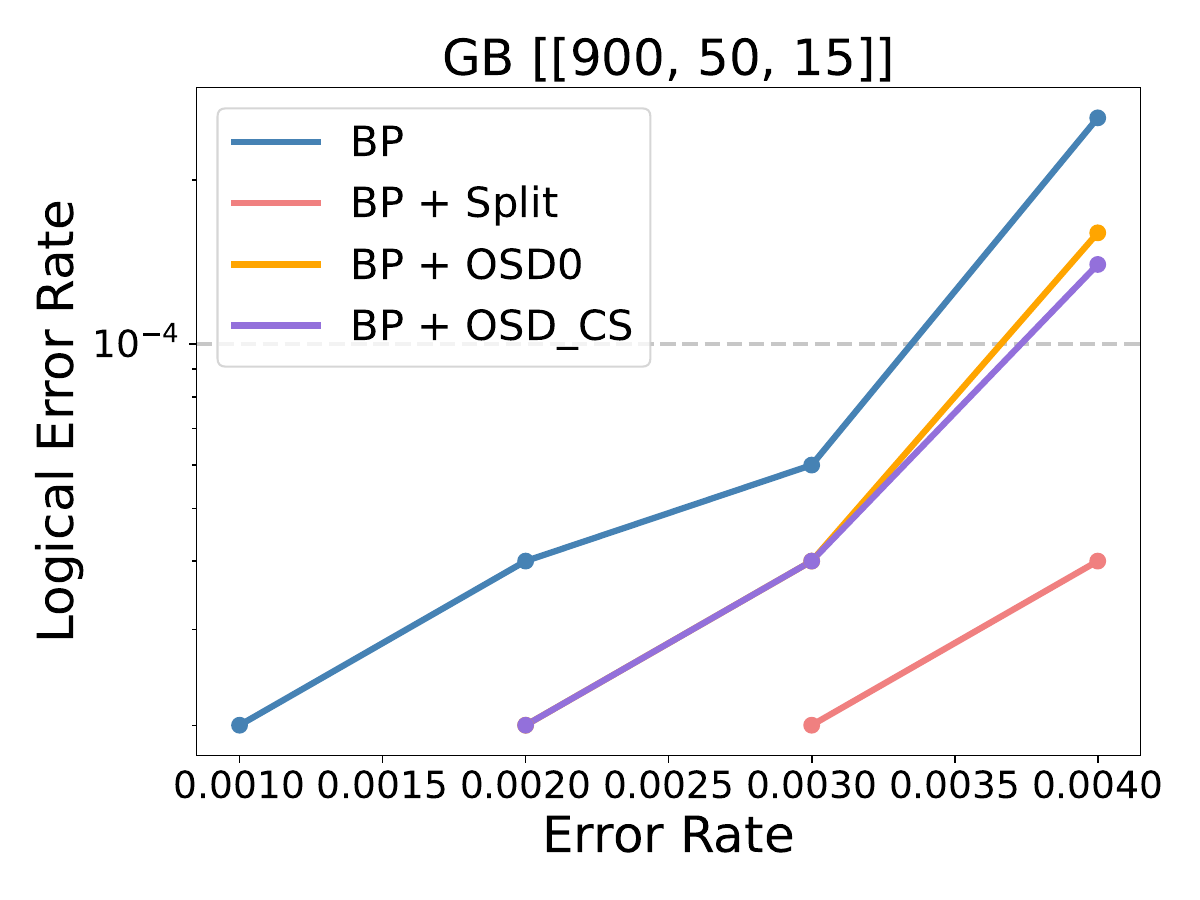} \\
        \hspace{-15pt}
        (a)\hspace{160pt}(b)\hspace{150pt}(c)
        \caption{Our Performance on Other QLDPC Codes: The notation \([[n, k, d]]\) represents a specific QEC code, where \(n\) denotes the number of data qubits, \(k\) indicates the number of encoded logical qubits, and \(d\) specifies the code distance. \textit{HP} refers to the hypergraph product code, while \textit{GB} refers to the generalized bicycle code.}
        \label{fig:general_code}
\end{figure*}

\subsection{Experimental Setup}
\label{sec:setup}

\noindent \textbf{Metrics.}
We evaluate our decoder’s performance using two metrics: \emph{logical error rate} (LER) and \emph{decoding time} ($T_{dcd}$). LER measures the probability of a decoding failure. A decoding failure occurs when the corrected logical state deviates from the initially encoded state. This metric provides a clear and effective measure of the decoder’s reliability in preserving quantum information under noise. The decoding time $T_{dcd}$ underscores our decoder’s time efficiency.

\vspace{3pt}
\noindent \textbf{Baselines.}
We evaluate \frameworkname~against the following baselines: (1) BP, (2) \emph{BP+OSD-0}, and (3) \emph{BP+OSD-CS}. Comparisons with BP highlight the improvements achieved through our syndrome split operations, while comparisons with the mainstream solutions (2) and (3) showcase the competitive performance of our decoder.


\vspace{3pt}
\noindent \textbf{Error Model.}
We evaluate our decoder using the widely recognized circuit-level error model~\cite{vittal2024flag, alavisamani2024promatch, bravyi2024high, fowler2009high}, which assumes each operation is either ideal with probability $1-p$ or faulty with probability $p$, where $p$ is the error rate. Faults occur independently across operations. A faulty CNOT is followed by one of $15$ non-identity Pauli errors applied uniformly to the control and target qubits. Faulty initialization prepares an orthogonal state, and faulty measurements introduce classical bit-flip errors. Idle qubit faults are modeled as random Pauli errors ($X, Y$, or $Z$). 

We choose the range of physical error rates for evaluation as $p = 10^{-3}$ to $5 \times 10^{-3}$. This range is well-justified for two key reasons: (1) It aligns with the typical noise levels observed in current quantum technologies~\cite{bravyi2022future, bluvstein2024logical, moses2023race, slussarenko2019photonic}, ensuring our evaluation is relevant to practical hardware. (2) This range lies just below the estimated error thresholds of the tested code instances, allowing us to observe meaningful error suppression behavior.

\vspace{3pt}
\noindent \textbf{Benchmarks. }
The benchmark includes six \emph{bivariate bicycle} (BB) codes, ranging from size \([[72, 12, 6]]\) with 72 qubits and encoding rate \(1/12\) to \([[784, 24, 24]]\) with 784 qubits and encoding rate \(1/95\), as well as larger codes from the \emph{hypergraph product} (HP) family, including \([[1922, 50, 16]]\) with 1922 qubits and encoding rate \(1/77\) and \([[7938, 578, 16]]\) with 7938 qubits and encoding rate \(1/28\). Additionally, we evaluate a popular \emph{generalized bicycle} (GB) code \([[900, 50, 15]]\) with 900 qubits and encoding rate \(1/18\). The diverse set of codes provides a comprehensive test for our decoder’s performance across a wide spectrum of realistic QEC scenarios.





\subsection{Main Results}\label{subsection: main results}

This section presents the main results of our evaluation, comparing \frameworkname~against three baseline approaches: (1) BP, (2) BP+OSD-0, and (3) BP+OSD-CS. The evaluation focuses on two key metrics: LER and decoding time ($T_{dcd}$).

\noindent \textbf{Comparison with BP.}
Overall, our decoder achieves an average 16.17× reduction in LER compared to BP on BB codes, driven by its effective handling of degeneracy through syndrome split. This advantage becomes more significant with larger code sizes, as shown in Fig.~\ref{fig:general_performance}(a)-(d) and Fig.~\ref{fig:general_performance}(e)(g). We attribute this to \frameworkname’s ability to address the degeneracy issue, which becomes more severe in larger codes, combined with its code-specific syndrome split operation that leverages the unique 2-layer structure and connectivity of BB codes, further boosting its performance over standard BP.

While achieving a lower LER, \frameworkname~introduces minimal decoding time overhead, as illustrated by the red (\frameworkname) and blue (BP) curves in Fig.~\ref{fig:general_performance}(f)(h). For instance, at a physical error rate of 0.003, the average time overhead is just 18.97\% and remains consistent across varying code sizes. This demonstrates that our syndrome split operation is computationally efficient, adding only a mild time overhead compared to BP.



\vspace{3pt}
\noindent \textbf{Comparison with BP+OSD. }
When compared with BP+OSD decoders, \frameworkname~demonstrates a significant improvement in LER, achieving an average reduction of 3.23x for BP+OSD-0. In small-scale codes, \frameworkname~achieves a comparable LER to BP+OSD-CS, while in large-scale codes, \frameworkname outperforms BP+OSD-CS significantly, achieving a much lower LER, as indicated by the red and purple curves in Figures~\ref{fig:general_performance}(e) and (g). This improvement arises from \frameworkname’s ability to adaptively modify the Tanner graph using syndrome split, consistently removing degeneracy
and altering the way BP updates error probabilities, ultimately yielding more reliable error estimations. In contrast, BP+OSD decoders simply run multiple BP iterations on a static graph,
leading to faulty error estimations that the OSD step may fail to correct. This advantage becomes more pronounced for larger code size, as shown in Fig.\ref{fig:general_code}(e)(g), where degeneracy issues become more severe in larger codes.
 
Additionally, \frameworkname~achieves a significant reduction in decoding time ($T_{dcd}$), as highlighted by the comparison of the orange (BP+OSD-0), purple (BP+OSD-CS), and red (\frameworkname) curves in Fig.~\ref{fig:general_code}(f)(h). This improvement in time efficiency becomes even more pronounced as the physical error rate increases. The advantage arises from \frameworkname’s lower time complexity compared to the OSD method, allowing it to sustain efficient decoding even as higher error rates increase computational demands.

\vspace{3pt}
\emph{Performance on Other qLDPC Codes}
As shown in Fig.~\ref{fig:general_code}, our decoder achieves significant LER improvements across three large-scale qLDPC codes (HP and GB codes), using the general syndrome split methods described in Sec.~\ref{subsec: syn split operation}, compared to both BP and BP+OSD decoders. This demonstrates that the general syndrome split operations introduced in Sec.~\ref{subsec: syn split operation} effectively address the degeneracy issue across a broad range of qLDPC codes.



\begin{figure*}[tp]
        \centering
        \includegraphics[width=0.32\linewidth]{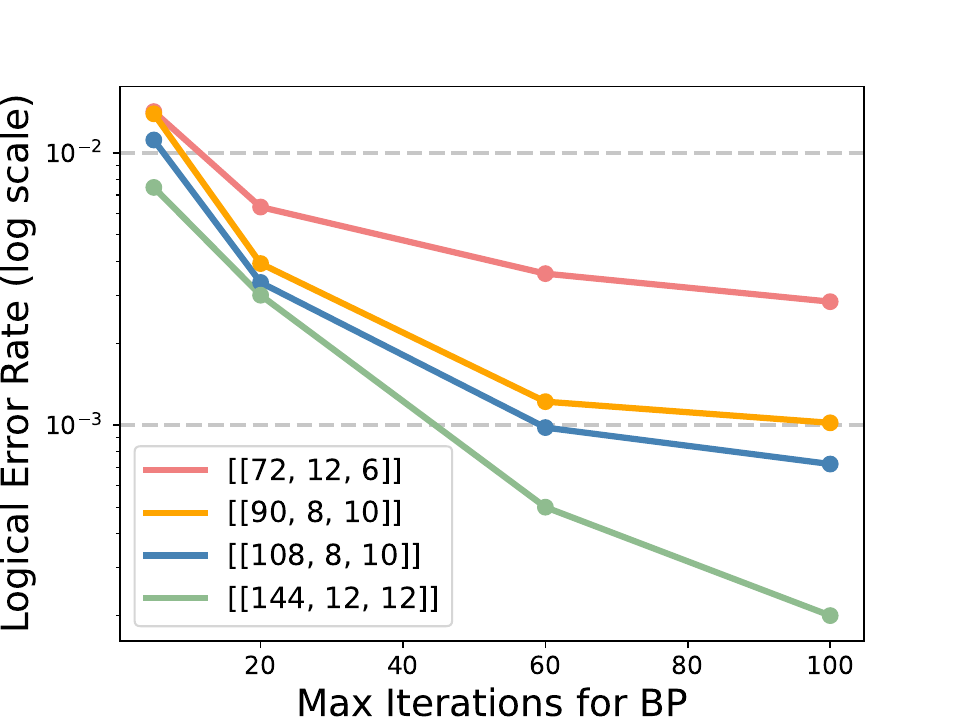}
        \includegraphics[width=0.32\linewidth]{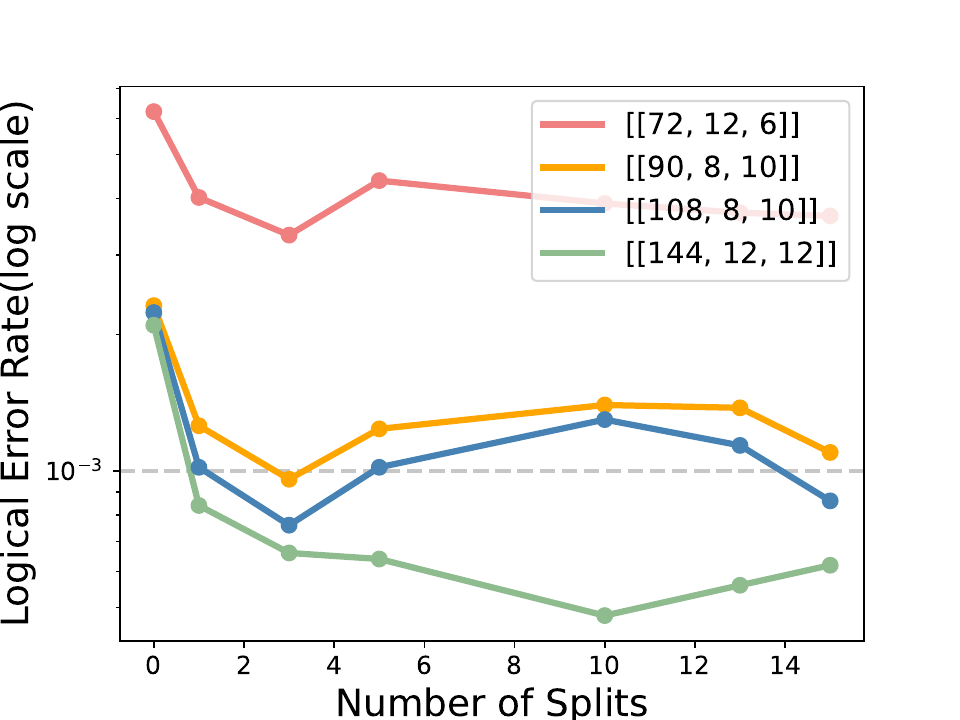} 
        \includegraphics[width=0.32\linewidth]{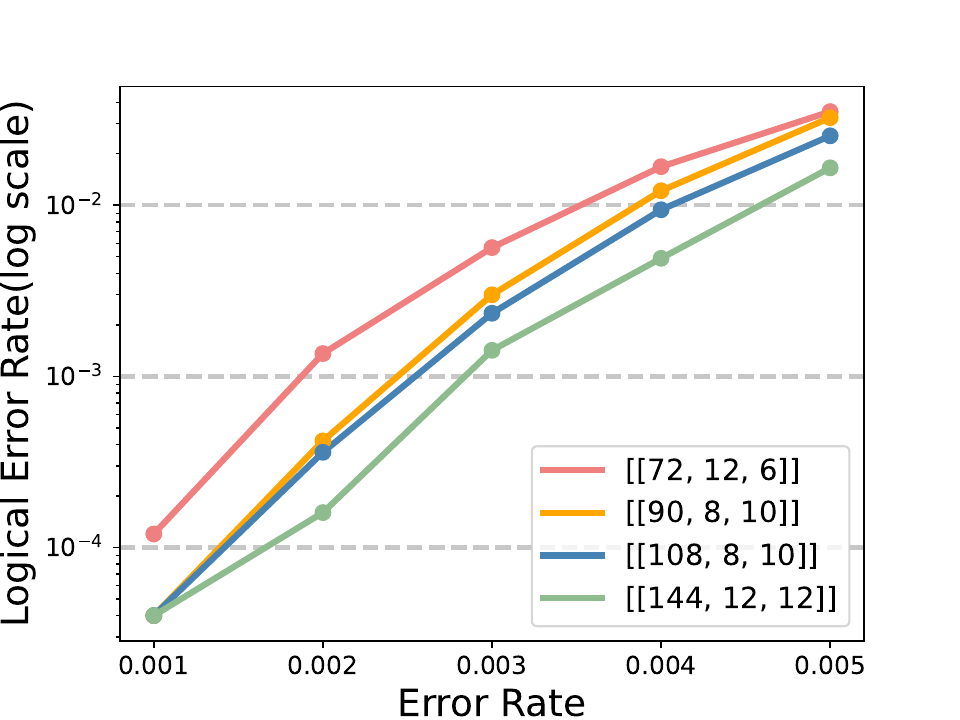} \\
        \hspace{-15pt}
        (a)\hspace{160pt}(b)\hspace{150pt}(c)
        \caption{Sensitivity Analysis: (a) illustrates how the LER varies with different given maximum iterations of BP, (b) examines the impact of the number of splits on the LER, and (c) presents the decoding LER across various physical error rates.}
        
        \label{fig:sensitivity}
\end{figure*}

\subsection{Sensitivity Analysis}\label{subsection: sensitivity}

In this subsection, we analyze the impact of three key factors on our decoder’s performance: the maximum BP iterations, number of splits, and physical error rate. Each parameter affects the decoder’s accuracy and efficiency, providing insights into its robustness and scalability under varied conditions.

\vspace{3pt}
\noindent \textbf{Max Iterations for BP}
As shown in Fig.~\ref{fig:sensitivity}(a), increasing the maximum number of BP iterations initially leads to a sharp decrease in logical error rate, followed by diminishing returns in improvement. However, more iterations also increase decoding time, making it essential to select an optimal number of iterations for each code to balance accuracy and efficiency. This highlights that simply increasing the number of BP iterations cannot consistently enhance decoding accuracy. To achieve better performance, occasional modifications to the Tanner graph are necessary to guide BP toward improved error estimations. Our syndrome split operation offers an effective solution to achieve this. 

\vspace{3pt}
\noindent \textbf{Number of Splits.}
As shown in Fig.~\ref{fig:sensitivity}(b), the LER initially decreases as the number of splits increases but starts to rise slightly beyond a certain threshold. This behavior arises because a moderate number of splits effectively resolves quantum degeneracy, while an excessive number introduces additional errors when assigning new syndromes to split checks, hindering subsequent BP rounds from converging to correct error estimations. These results highlight the importance of carefully balancing the number of splits with BP iterations to achieve optimal performance.

\vspace{3pt}
\noindent \textbf{Physical Error Rate.}
As shown in Fig. \ref{fig:sensitivity}(c), the logical error rate increases as the physical error rate rises. This trend occurs because higher physical error rates result in more complex error syndromes, making it more challenging for the decoder to assign accurate syndromes to the split checks and leading to more decoding failures.

\begin{figure*}[tp]
        \centering
        \includegraphics[width=0.325\linewidth]{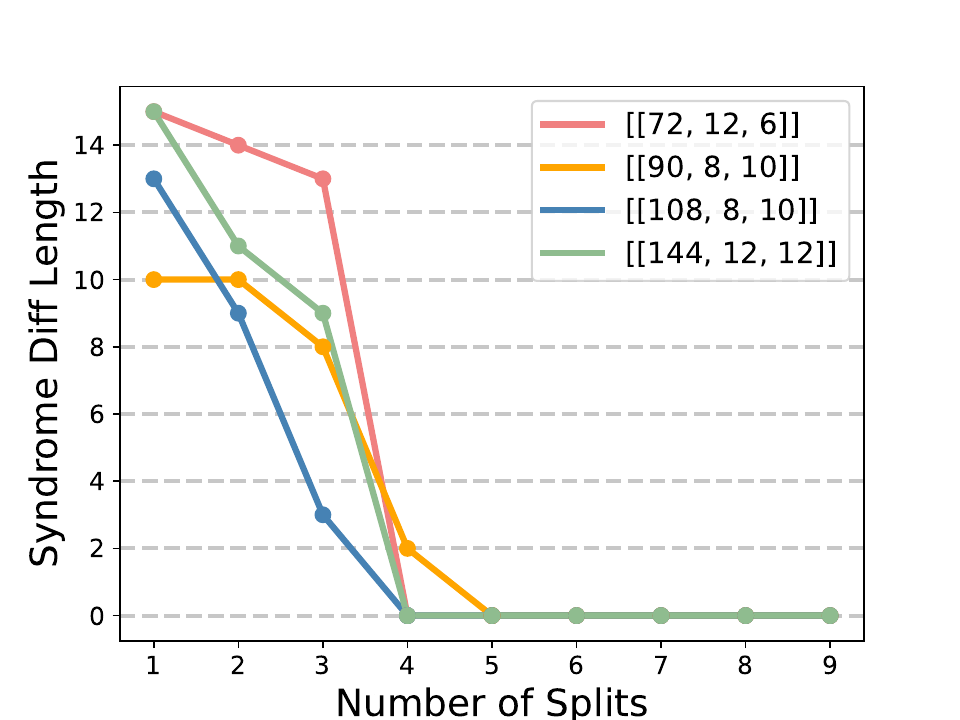}
        \includegraphics[width=0.325\linewidth]{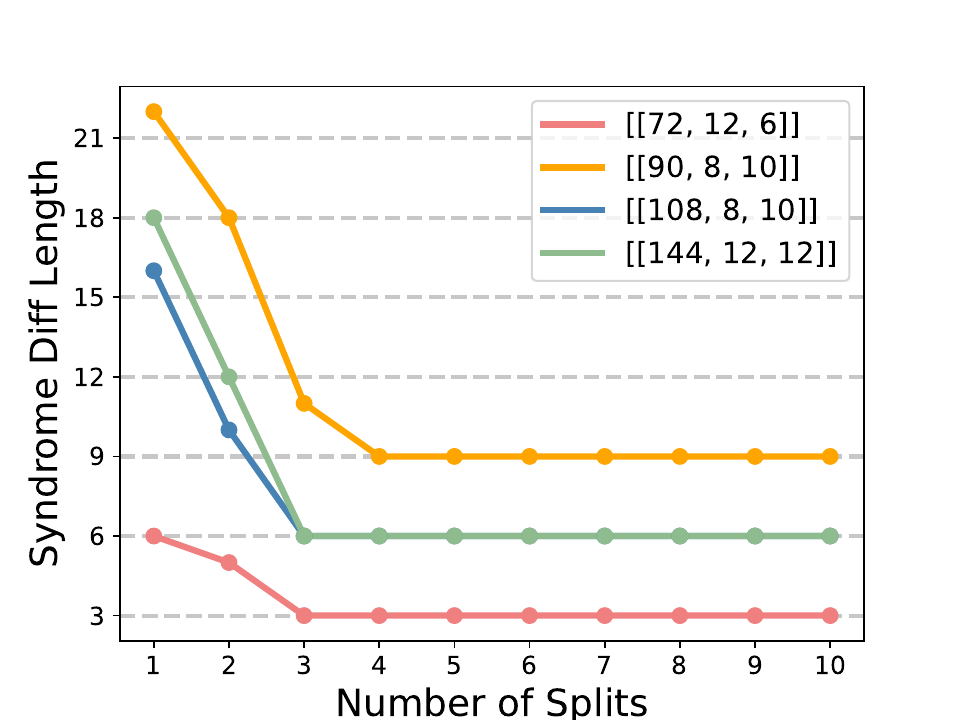} 
        \includegraphics[width=0.325\linewidth]{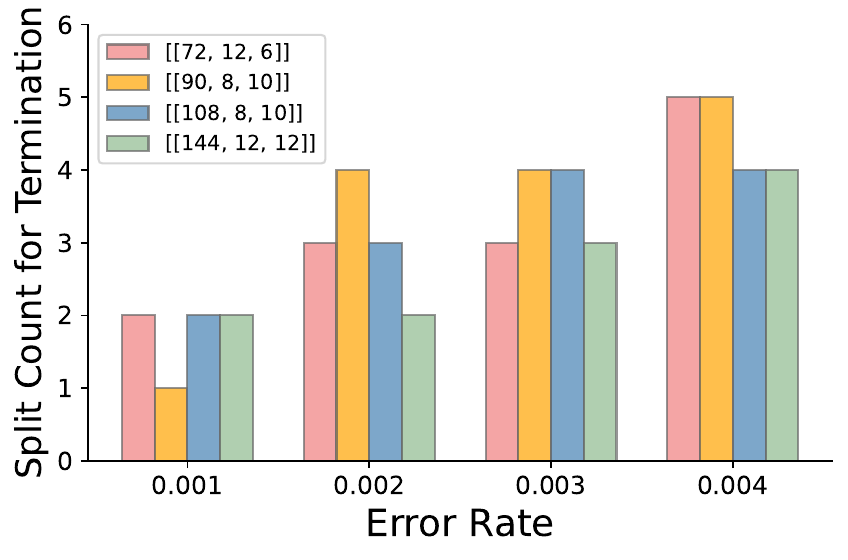} \\
        
        (a)\hspace{160pt}(b)\hspace{150pt}(c)
        \hspace{-15pt}
        \caption{Split Termination Process: (a) shows the scenarios of successful decoding as the number of splits increases, (b) illustrates the scenarios of unsuccessful decoding with an increasing number of splits, and (c) presents the number of effective splits used before termination.}
        
        \label{fig:tuned_split_num}
\end{figure*}

\subsection{Auto-Termination for Split Optimization}\label{subsection:termination}

To enhance the efficiency of our decoder, we propose an auto-termination mechanism that dynamically determines the optimal point to stop the splitting process. This mechanism monitors the behavior of the \emph{Syndrome Difference Length}, defined as the Hamming distance between corrected syndrome and the original input syndrome at each split step ($|H\hat{e}-s|$ in Sec.~\ref{sec: check split}). The syndrome difference length provides a quantitative measure of the decoder's progress, depicting how closely the corrected syndrome matches the original syndrome.

Fig.~\ref{fig:tuned_split_num}(a) shows the \emph{successful convergence case for \frameworkname} and Fig.~\ref{fig:tuned_split_num}(b) shows the \emph{unsuccessful convergence cases}. We observe that the syndrome difference length exhibits distinct convergence patterns during the splitting process. In successful cases, the syndrome difference length gradually decreases and eventually reaches zero, indicating convergence to a valid solution. In unsuccessful cases, the syndrome difference length stabilizes at a non-zero value, signifying that further splitting will not result in convergence.

By dynamically tracking the syndrome difference length at each split step, the auto-termination mechanism identifies stabilization, either at zero (successful decoding) or at a constant non-zero value (unsuccessful decoding). Once stabilization is detected, the mechanism terminates the splitting process, thereby reducing unnecessary further splits and BP iterations.


Additionally, Fig.~\ref{fig:tuned_split_num}(c) examines how the \emph{Split Count for Termination} varies with different physical error rates. The results show that higher error rates require more splits, reflecting the increased decoding complexity. This observation underscores the adaptability of the proposed mechanism to varying error conditions, ensuring both effectiveness and efficiency in the decoding process.

\section{Related Work}
This section provides a broader overview on qLDPC decoders.

\vspace{3pt}
\noindent\textbf{Matching-Based Decoders.} These decoders reformulate the decoding problem into the MWPM problem in graph theory. They are primarily designed for surface codes~\cite{bravyi1998quantum, dennis2002topological, fowler2012surface}, achieving significant advancements over the past two decades. Current approaches reach $1\%$-$10\%$  error thresholds across various settings~\cite{demartireview, higgott2022pymatching, higgott2023sparse}. The complexity has been reduced from the original near-cubic runtime~\cite{edmonds1965paths} to linear time~\cite{wu2023fusion}, with support for parallelism~\cite{fowler2013minimum} and hardware-efficient FPGA implementations~\cite{alavisamani2024promatch, vittal2023astrea}. Despite their effectiveness for surface codes, they cannot be directly applied to more general qLDPC codes with higher qubit connectivity.

\vspace{3pt}
\noindent\textbf{BP-Based Decoders.} BP is an effective decoder for classical LDPC codes~\cite{mackay1997near}, but its effectiveness diminishes for qLDPC codes due to degeneracy. Various post-processing techniques have been proposed to address this, including order statistics decoding (OSD)~\cite{panteleev2021degenerate, roffe2020decoding}, ambiguity clustering~\cite{wolanski2024ambiguity}, and localized statistics decoding~\cite{hillmann2024localized}. While these methods achieve high accuracy, they often involve matrix inversion, significantly increasing decoding time. Alternative approaches that combine BP with techniques such as stabilizer inactivation~\cite{du2022stabilizer}, guided decimation~\cite{yao2024belief}, and closed-branch decoding~\cite{demarti2024closed} reduce time complexity but offer limited accuracy.

\vspace{3pt}
\noindent\textbf{Other Decoders.} The small-set flip (SSF) decoder~\cite{leverrier2015quantum, fawzi2018efficient, fawzi2020constant} was developed for hypergraph product codes~\cite{tillich2013quantum}, a key family of qLDPC codes. It achieves a high threshold of $7.5\%$ in numerical experiments~\cite{grospellier2021combining, grospellier2018numerical}, but its effectiveness is limited for other qLDPC code families. Additional qLDPC decoders include union-find decoders~\cite{delfosse2022toward}, layered decoders~\cite{du2023layered}, and decoders inspired by classical LDPC algorithms~\cite{quintavalle2022reshape}.
\section{Conclusion}
We introduced~\frameworkname, a novel decoder that adaptively modifies the decoding graph to overcome the limitations of decoding of qLDPC codes. By addressing quantum degeneracy—one of the primary challenges in BP—our approach achieves both high accuracy and low latency, making it a viable solution for real-time QEC. The core of \frameworkname~is the innovative syndrome split operation, which carefully disrupts degeneracy structures while preserving error information. By leveraging BP’s intermediate outputs, the decoder dynamically identifies syndrome nodes likely affected by degeneracy and splits them in a manner that guides BP to converge toward accurate error estimations. This adaptive approach not only enhances BP performance but also eliminates the need for time-intensive post-processing OSD steps. We believe our syndrome split idea could open up a promising new direction for further research.

\section{Acknowledgments}
We thank the anonymous reviewers for their constructive feedback. This work is supported in part by NSF 2048144, NSF 2422169, NSF 2427109. This material is based upon work supported by the U.S. Department of Energy, Office of Science, National Quantum Information Science Research Centers, Quantum Science Center (QSC). This research used resources of the Oak Ridge Leadership Computing Facility, which is a DOE Office of Science User Facility supported under Contract DE-AC05-00OR22725. This research used resources of the National Energy Research Scientific Computing Center (NERSC), a U.S. Department of Energy Office of Science User Facility located at Lawrence Berkeley National Laboratory, operated under Contract No. DE-AC02-05CH11231. We thank Andrew Cross for the technical discussions and feedback throughout the project. 

\bibliographystyle{unsrt}
\bibliography{references}

\end{document}